\documentclass{elsarticle}
\usepackage{amssymb, amsmath, bm, graphicx, adjustbox, lscape,breakcites}
\usepackage[margin=2.5cm]{caption}
\usepackage[margin=0.8in]{geometry}
\usepackage[table,xcdraw]{xcolor}
\usepackage{soul, natbib}
\sethlcolor{green}
\usepackage{subfigure, epstopdf, siunitx, ifthen, algpseudocode, algorithm}
\usepackage{booktabs,tikz,xtab, multirow}
\PassOptionsToPackage{hyphens}{url}\usepackage[breaklinks=true]{hyperref}
\hypersetup{colorlinks = true,linkcolor = blue,anchorcolor =red,citecolor = blue,filecolor = red,urlcolor = red, pdfauthor=author}
\usepackage[ruled,english,algo2e,algoruled,vlined,linesnumbered]{algorithm2e}
\usepackage{perpage} 
\MakePerPage{footnote} 

\makeatletter
\def\ps@pprintTitle{%
 \let\@oddhead\@empty
 \let\@evenhead\@empty
 \def\@oddfoot{}%
 \let\@evenfoot\@oddfoot}
\makeatother

\graphicspath{{figs/}}

\begin{document}
\begin{frontmatter}
\title{Deep Graph Convolutional Reinforcement Learning for Financial Portfolio Management - DeepPocket}
\author{Farzan Soleymani}
\ead{Farzan.Soleymani@nrc-cnrc.gc.ca}
\author{Eric Paquet\corref{cor1}}
\ead{Eric.Paquet@nrc-cnrc.gc.ca}
\cortext[cor1]{Corresponding author}
\address{National Research Council, 1200 Montreal Road, Ottawa, ON K1K 2E1, Canada}
\begin{abstract}
Portfolio management aims at maximizing the return on investment while minimizing risk by continuously reallocating the assets forming the portfolio. These assets are not independent but correlated during a short time period. A graph convolutional reinforcement learning framework called DeepPocket is proposed whose objective is to exploit the time-varying interrelations between financial instruments. These interrelations are represented by a graph whose nodes correspond to the financial instruments while the edges correspond to a pair-wise correlation function in between assets. DeepPocket consists of a restricted, stacked autoencoder for feature extraction, a convolutional network to collect underlying local information shared among financial instruments and an actor–critic reinforcement learning agent. The actor--critic structure contains two convolutional networks in which the actor learns and enforces an investment policy which is, in turn, evaluated by the critic in order to determine the best course of action by constantly reallocating the various portfolio assets to optimize the expected return on investment. The agent is initially trained offline with online stochastic batching on historical data. As new data become available, it is trained online with a passive concept drift approach to handle unexpected changes in their distributions. DeepPocket is evaluated against five real-life datasets over three distinct investment periods, including during the Covid-19 crisis, and clearly outperformed market indexes.
\end{abstract}
\end{frontmatter}
\section{Introduction}

The process of selecting and managing a group of financial instruments such as stocks, bonds, and securities, is called portfolio management \citep{markowitz1978portfolio}. Portfolio management aims at maximizing the return on investment while minimizing the risk. Stock market movements vary with time and reflect social and political trends. As a result, the financial market follows a complex trajectory that is characterized by its volatility as well as by the correlations between financial instruments. One must have a broad knowledge of these movements and erratic behaviors in order to somehow predict market evolution.  Various strategies for portfolio management have been proposed in the literature such as \citep{pouya2016solving} which employs a multiobjective weed optimization method, \citep{ yue2017new} which relies on fuzzy multi-objective, high-order moment portfolio selection, and \citep{omidi2017efficient} which dynamically solves portfolio selection with uncertainty chance constraints.  Constrained portfolio selection has been discussed in \citep{garcia2019selecting, garcia2019credibilistic}. In this paper, it is proposed to learn the portfolio allocation directly from the data with a deep neural network: a free-form, data-driven approach.

Investing in multiple indexes and stocks could be overwhelming as one must assess market evolution while continuously reallocating assets in the most profitable way. Market movements, price fluctuations, and sudden turmoil are known to be interrelated \citep{park2013stock}. Yet, it is extremely difficult to model the duration and magnitude of these fluctuations \citep{sarwar2019interrelations, park2013stock}. For instance, the financial market is influenced by exogenous factors related to the globalized economy, such as the merging of $21^{st}$ Century Fox and Disney, or Peugeot Fiat-Chrysler. Understanding the dependencies and interrelationships among assets helps to understand market evolution while identifying the optimal portfolio composition and risk management strategy \citep{drozdz2001towards, elton2009modern}. Deep learning techniques are well suited for extracting complex patterns from large datasets \citep{bengio2015deep}. In addition, datasets are mostly Euclidean in the sense that the space associated with the data has a Euclidean geometry.

Images are a good example of such a geometry as the pixels form a regular grid which spans a bidimensional Euclidean space. An important benefit of Euclidean space is that it permits the concept of convolution. This is paramount for convolutional networks as their architecture is based on this very concept. Convolutional neural networks (CNNs) have been widely employed for pattern recognition, localization, and parameter reduction on Euclidean datasets, just to mention a few. If the space is not Euclidean, convolution networks cannot be employed in their standard form \citep{defferrard2016convolutional}. Moreover, most deep learning algorithms assume that the instances forming the dataset are statistically independent \citep{gama2013evaluating, barabasi2016network, wu2019comprehensive}; but this assumption does not apply to datasets whose instances are interrelated. Still, non-Euclidean data are far from rare. One may think, for instance, of scientific publications and citations\citep{defferrard2016convolutional, velivckovic2017graph} in which the various instances are linked to each other, thereby forming a non-Euclidean interconnected network. The same may be said about macromolecular structures, whose importance in medicine cannot be overstressed \citep{gilmer2017neural}.

Financial data, such as financial instruments, are no exception as they also define a non-Euclidean space \citep{wu2020comprehensive}. Indeed, the reason lies within the interrelation between financial instruments \citep{lucey2011robust}. These are correlated \citep{anthony1988interrelation} during short periods of time, especially in the presence of exogenous extreme events \citep{ arfaoui2017oil}. Their structure may be represented by a graph in which the nodes correspond to the financial instruments, while the edges or links correspond to some pair-wise correlation function. Because the geometry is non-Euclidean, one cannot easily transpose the Euclidean definitions of locality, translation, and compositionality which underpin convolution and, ultimately, are needed to employ convolutional neural networks \citep{angles2008survey, bronstein2017geometric}. In Euclidean geometry, convolution may be performed directly or with the help of the convolution theorem \citep{hammond2011wavelets} by employing the Fourier transform. The latter is well defined in Euclidean space.

In order to apply the convolution theorem in finance, one must define a Fourier transform directly on the non-Euclidean graph. Such a graph Fourier transform was introduced recently by \citep{shuman2013emerging}: it allows convolutions on the graph to be evaluated directly with the help of the convolution theorem and the graph Fourier transform. Once defined, the transform makes it possible to apply a convolutional network directly to non-Euclidean graphs, which are henceforth known as graph convolutional networks (GCN) \citep{shuman2013emerging, henaff2015deep}. As for their Euclidean counterparts, they consist of a bank of kernels (or filters) which are convolved with their input data. Euclidean filters are local in the sense that their extension (dimensionality) is much smaller than that of the signals with which they are convolved. In the non-Euclidean case, because of the Fourier transform, localization is lost, generally speaking. This problem may be overcome if the filters are defined with truncated Chebyshev polynomials, which are characterized by their compact support and low computational complexity (recurrence relation) \citep{defferrard2016convolutional}. 

The portfolio management problem is inherently nonlinear, stochastic, and time-dependent. Therefore, it may be portrayed as a decision-making process that yields a sequence of actions. These actions are defined as the amount of funds to be allocated to each asset forming the portfolio to increase the expected return on investment. This is achieved by training an agent which selects a set of actions from all possible actions in the action space, aiming to increase the portfolio return over a certain investment period. This objective may be formulated as a stochastic optimization problem such that the solution yields an optimal chain of actions which aims at maximizing the expected return on investment. 

These types of problem can be solved using reinforcement learning (RL) techniques where an agent takes action $(\mathbf{a}_{t})$ from an action space $(\mathcal{A})$ based on the state $(\mathbf{s}_{t})$ of the environment which belongs to a state-space $(\mathcal{S})$. Each action can be interpreted as a kind of interaction with the environment, which is associated with a scalar reward $r_{t}$. RL determines an optimal path in the action space, which aims at increasing the reward. The path is characterized by a chain of actions that are derived from a policy function $\pi(\mathbf{a}_{t}|\mathbf{s}_{t})$. Therefore, the agent learns a policy based on past experience, which is evaluated by a value function according to some environment dynamics \citep{andrew1999reinforcement}. Since the portfolio management problem is time-dependent, learning must be conducted online.

There are various RL techniques that provide optimal policy and value functions, e.g. Q-learning \citep{watkins1992q, bu2008comprehensive}, SARSA \citep{zhao2016deep}, deep deterministic policy gradient (DDPG) \citep{lillicrap2015continuous} and actor--critic \citep{konda2000actor} just to mention a few. Q-learning is an off-policy algorithm meaning that the policy that generates the actions may be unrelated to the improved policy. On the other hand, SARSA is an on-policy approach that aims to improve the policy that dictates the actions \citep{andrew1999reinforcement}. DDPG is a model-free, off-policy approach that learns a deterministic policy in a continuous space \citep{lillicrap2015continuous}. The advantage of such a policy is its ability to explore all possible actions while learning a deterministic policy \citep{andrew1999reinforcement}. Nonetheless, this exploration is time-consuming as it requires a large number of training epochs. In the DeepBreath framework, a SARSA algorithm was employed in order to learn the optimal policy \citep{soleymani2020financial}. However, SARSA may lead to abrupt changes in the policy which are not necessarily suitable for portfolio optimization \citep{celikyurt2007multiperiod} as a result of an erratic change in the policy if the latter performs too poorly \citep{andrew1999reinforcement}.

The actor--critic algorithm consists of two components: one being associated with the policy while the other is associated with the value function. The network responsible for the implementation of the policy is known as the actor. The critic is a learnable network that evaluates the performance index (value function) for each new each state based on the actions previously taken by the actor. Therefore, the actor--critic approach may be implemented with two online, trainable neural networks that work concurrently in tandem. In this work, the actor--critic algorithm is exploited wherein the actor is trained based on experience, corresponding to probabilistic mapping from state actions according to a learnable policy. At the same time, the critic estimates the expected future return through a state-value function while constantly improving the policy \citep{bhatnagar2009natural}. Both the actor and the critic are implemented with specially designed convolutional networks where the output of the actor determines the portfolio allocation while the return on investment is approximated by the critic. One of the most important innovations in this work is the introduction of a \textbf{graph convolutional network} for the actor and the critic in order to take advantage of the correlation (as well as the non-Euclidean space) existing in between the financial instruments.

A graph convolutional reinforcement learning framework, DeepPocket is introduced, consisting of a restricted, stacked autoencoder (RSAE) for feature extraction and dimensionality reduction, a graph convolutional network (GCN) to acquire interrelations among financial instruments, as well as a convolutional network for each of the actor and the critic. Feature extraction generates a low-dimensional and information-rich description which is more suited to machine learning, besides reducing the computational complexity. The GCN captures the correlations existing between financial instruments. The investment policy is enforced, and the return on investment is estimated using the actor--critic algorithm. The framework is initially trained offline; then, the weights of the neural networks are updated online as new data becomes available. The offline training step is conducted using historical data, while online training is based on passive concept-drift detection \citep{gama2014survey}.

The paper is organized as follows. The mathematical model of portfolio management is discussed in Section~\ref{sec:math_model}. The architecture of the proposed framework is presented in Section~\ref{sec:system_architecture}, which includes feature normalization in subsection~\ref{subsec:feature_normal}, the restricted stacked autoencoder in subsection~\ref{subsec:RSAE}, and the graph convolutional networks in subsection~\ref{subsec:Spect_GNN}. The reinforcement learning framework is introduced in Section~\ref{sec:Deep_RL}. This is followed by a description of our offline and online learning approaches in Section~\ref{sec:online_learning}. Our experimental results are presented in Section~\ref{sec:EXPresults}. Finally, Section~\ref{sec:Discussion} concludes the paper.

\section{Mathematical Model}\label{sec:math_model}
\begin{figure}[hbt!]
  \centering
  \includegraphics[width=0.65\textwidth]{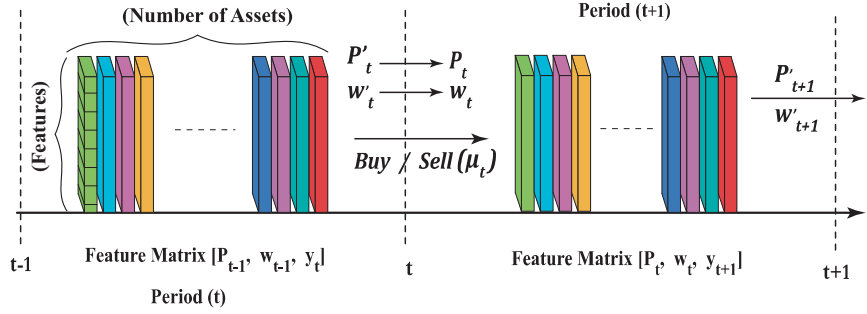}
  \caption{Weight reallocation process. The portfolio weight vector $\mathit{\mathbf{w}'_{t-1}}$ and the portfolio value $\mathit{P'_{t-1}}$ at the beginning of period $(t)$ evolve to ($\mathit{t}$) and $\mathit{P'_{t}}$ respectively. At that moment, assets are sold and bought in order to increase the expected return on investment. These operations involve commission fees which shrink the portfolio value to $\mathit{P_{t}}$ which result in new weights $\mathit{\mathbf{w}_{t}}$.}
  \label{fig:trading_window}
\end{figure}

A portfolio containing $m$ assets may be managed directly by an individual or by financial professionals through an institution like an investment bank. The investors must select and allocate funds to a group of assets according to an investment strategy, subject to some risk aversion. 
Our portfolio management model is inspired by an approach introduced by \citep{ormos2013performance} which introduces an experimental approximation of the log-optimal investment strategy which ensures a near optimal growth rate of investments along with a survivorship bias-free setup. In this model, a trading period corresponds to one business day. The closing price of all assets in the portfolio corresponds to the vector $V_{t}$. The vector also contains the amount of available cash. The normalized price vector is defined as
\begin{equation}\label{eq:price_relative_vector}
  \mathbf{Y}_{t} := \mathbf{V}_{t} \oslash \mathbf{V}_{t-1} = \left[1,\frac{\mathbf{V}_{1,t}}{\mathbf{V}_{1,t-1}}, \frac{\mathbf{V}_{2,t}}{\mathbf{V}_{2,t-1}},\cdots,\frac{\mathbf{V}_{m,t}}{\mathbf{V}_{m,t-1}}\right]^{T}.
\end{equation}
where $(\oslash)$ denotes element-wise division. Initially, there is no assumption regarding the weight distribution: as a result, the portfolio only consists of cash which means that the weight associated with the cash is one (1), while the other weights are zero (0):
\begin{equation}\label{eq:init_weight}
  \mathbf{w}_{0} = [1,0,\cdots,0]^{T}.
\end{equation} 
A transaction factor, which shrinks the portfolio value, is associated with each transaction, such as buying or selling. At the end of a business day, as illustrated in Fig.~\ref{fig:trading_window}, the portfolio weights vector is equal to:
\begin{equation}\label{eq:ptf_weight}
  \mathbf{w}'_{t} = \frac{\mathbf{Y}_{t}\odot \mathbf{w}_{t-1}}{\mathbf{Y}_{t}\cdot \mathbf{w}_{t-1}},
\end{equation}
At the beginning of the period $(t)$, the portfolio weight vector is $\mathbf{w}_{t-1}$. The agent enforces the investment policy in order to reallocate the assets, such as to increase the expected return on investment, resulting, at the end of the trading day, in a new weight vector $\mathbf{w}'_{t}$. A commission fee $\mu \in (0, 1]$ is applied to the transactions, resulting in the final weight vector $\mathbf{w}_{t}$.

\begin{equation}\label{eq:ptf_value}
  P_{t} = \mu_{t} P'_{t}.
\end{equation}
The portfolio value at the end of period $(t)$ is obtained by 
\begin{equation}
  P_{t} = \mu_{t} P_{t-1}\cdot\mathbf{Y}_{t}\cdot \mathbf{w}_{t-1}.
\end{equation}
while the return rate and logarithmic return rate for period $(t)$ are given by 
\begin{subequations}\label{return_rate}
  \begin{align}
  \label{eq:ret_rate}
  \rho_{t}&:= \frac{P_{t}}{P_{t-1}}-1 = \frac{\mu_{t}\cdot P'_{t}}{P_{t-1}}-1 = \mu_{t} \mathbf{Y}_{t}\cdot \mathbf{w}_{t-1}-1,\\
  \label{eq:log_return}
  R_{t}&:=\ln\frac{P_{t}}{P_{t-1}} = \ln(\mu_{t} \mathbf{Y}_{t} \mathbf{w}_{t-1})-1.
  \end{align}
\end{subequations}
Accordingly, the final portfolio value at time horizon $t_{f}$ is determined by
\begin{equation}\label{eq:ptf_final}
  P_{f} = P_{0} \cdot \exp{\left(\sum_{t=1}^{t_{f}+1} R_{t}\right)} = P_{0} \prod_{t=1}^{t_{f}+1}\mu_{t} \mathbf{Y}_{t} \mathbf{w}_{t-1}.
\end{equation}
Buying and selling assets result in spending and gaining cash \citep{soleymani2020financial}, which implies that:
\begin{align}\label{eq:sell_buy_cash}
  &\qquad \qquad \qquad \qquad C_{g} = (1-c_{s}))P'_{t}\sum_{i=1}^{m}\cdot ReLU( \mathbf{w'}_{t,i}-\mu_{t} \mathbf{w}_{t,i}) \\
  C_{s} &= (1-c_{b})\left[\mathbf{w'}_{t,0}+(1-c_{s})\sum_{i=1}^{m} ReLU(\mathbf{w'}_{t,i}-\mu_{t} \mathbf{w}_{t,i}) - \mu_{t} \mathbf{w}_{t,0}\right] = \sum_{i=1}^{m} ReLU(\mu_{t} \mathbf{w}_{t,i} - \mathbf{w'}_{t,i})
\end{align}
where the selling commission rate belongs to $0<c_{s}<1$, the buying commission rate lies within $0<c_{b}<1$ and $\mathop\textrm{ReLU}(x) = \max(0,x)$ is a rectified linear unit. As a result, the available cash $P’_{t}\mathbf{w}’_{t,0}$ becomes $\mu_{t}P’_{t}\mathbf{w}’_{t,0}$.
Therefore, the transaction factor $\mu_{t}$ is obtained by solving an implicit equation:
\begin{subequations}\label{eq:transaction_factor}
  \begin{align}
    \mu_{t} &= \frac{1}{1-c_{b}\mathbf{w}_{t,0}}\left[1-c_{b}\mathbf{w'}_{t,0} - (c_{s}+c_{b}-c_{s}c_{b})\sum_{i=1}^{m} ReLU(\mathbf{w'}_{t,i}-\mu_{t}\mathbf{w}_{t,i})\right] \\
    & \qquad \qquad \qquad \qquad \qquad \mu_{t} = \mu_{t}(\mathbf{w}_{t-1},\mathbf{w}_{t},\mathbf{Y}_{t})
  \end{align}
\end{subequations}
Eq.~\ref{eq:transaction_factor} cannot be solved analytically. A solution may, however, be achieved iteratively, convergence being guaranteed by a theorem demonstrated in \citep{jiang2017deep}. Our mathematical model relies on two assumptions: 
\begin{itemize}
  \item Trading is possible at any time during business hours.
  \item The volume of financial instruments traded by the agent is relatively small compared to the total number of assets traded during a day period. 
\end{itemize} 
As long as the volume of financial instruments traded by the agent is small compared to the overall traded volume, the latter assumptions remain valid. The architecture of the proposed framework is explained in the following section.
\section{System Architecture}\label{sec:system_architecture}
A profitable trading system must be able to forecast price movements on the stock market. These price movements may be characterized by various financial indicators such as price trend \citep{nti2019systematic}, the momentum indicator, and moving average, to mention a few (see Table~\ref{tab:Financial_Indicators}).

The price trend indicates the direction toward which the market evolves \citep{qiu2016predicting}. The strength of a trend, as well as the likeliness of its reversal, are measured by the momentum indicator, this being related to the rate at which prices change \citep{cervello2020forecasting}. The moving average indicator measures the average price of a given financial instrument over a certain period of time. Seven financial indicators were selected in order to measure price movement and predict future trends. These indicators include the opening and closing price of the market and the lowest, and the highest prices reached during the trading period, along with financial indicators such as the Hull moving average, average true range, the dynamic momentum index, the commodity selection index, among others \citep{murphy1999technical}. These indicators are described in Table~\ref{tab:Financial_Indicators}. 

\begin{figure}[hbt!]
  \centering
  \includegraphics[width=0.65\textwidth]{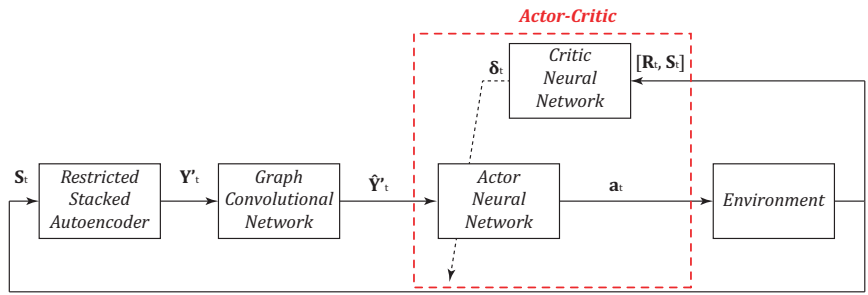}
  \caption{Global architecture of DeepPocket.}
  \label{fig:System_architecture}
\end{figure}

The proposed portfolio management framework consists of multiple components, namely the normalization module, feature selection with a restricted, stacked autoencoder, and a graph convolutional neural network which extracts information-rich features based on the correlations between financial instruments. The policy is learned with an actor--critic algorithm: the latter comprising two convolutional neural networks as illustrated in Fig.~\ref{fig:System_architecture}. These neural networks are fully described in the next section.
 
\subsection{Feature Normalization}\label{subsec:feature_normal}
Each asset is characterized by a feature vector containing twelve (12) features including opening, closing, low and high prices in addition to financial indicators such as average true range that evaluate the market volatility over a certain period \citep{soleymani2020financial} as illustrated in Table~\ref{tab:Financial_Indicators}. 

\begin{table}[H]
\caption{Financial Indicators Employed as Features.}
\label{tab:Financial_Indicators}
\small
\centering
\begin{tabular}{|c|c|}
\hline
\textbf{Financial Indicators} & \textbf{Definition} \\ \hline
\textit{Average True Range} & \begin{tabular}[c]{@{}c@{}}The average true range is a technical indicator that assess\\ the volatility of an asset in the market through a certain period.\end{tabular} \\ \hline
\textit{\begin{tabular}[c]{@{}c@{}}Commodity Channel\\ Index\end{tabular}} & \begin{tabular}[c]{@{}c@{}}The commodity channel index is a momentum-based\\ oscillator used to help determine cyclical trend of an asset price\\ as well as strength and direction of that trend.\end{tabular} \\ \hline
\textit{\begin{tabular}[c]{@{}c@{}}Commodity Selection\\ Index\end{tabular}} & \begin{tabular}[c]{@{}c@{}}The commodity selection index is a momentum indicator \\ that evaluates the eligibility of an asset for short term investment.\end{tabular} \\ \hline
\textit{Demand Index} & \begin{tabular}[c]{@{}c@{}}The demand index is an indicator that\\  uses price and volume to assess buying and\\ selling pressure affecting a security.\end{tabular} \\ \hline
\textit{\begin{tabular}[c]{@{}c@{}}Dynamic Momentum\\ Index\end{tabular}} & \begin{tabular}[c]{@{}c@{}}The dynamic momentum index determines \\ if an asset is an asset is overbought or oversold.\end{tabular} \\ \hline
\textit{\begin{tabular}[c]{@{}c@{}}Exponential\\ Moving Average\end{tabular}} & \begin{tabular}[c]{@{}c@{}}An exponential moving average is a technical indicator that\\ follow the price of an asset while prioritizing on recent data points.\end{tabular} \\ \hline
\textit{\begin{tabular}[c]{@{}c@{}}Hull\\ Moving Average\end{tabular}} & \begin{tabular}[c]{@{}c@{}}The hull moving average is a more responsive alternative to\\ moving average indicator that focus on the current price\\ activity whilst maintaining curve smoothness.\end{tabular} \\ \hline
\textit{Momentum} & \begin{tabular}[c]{@{}c@{}}The momentum in a technical indicator that determines the speed \\ at which the price of an asset is changing.\end{tabular} \\ \hline
\end{tabular}
\end{table}

Since the effective range and extrema of the features are unknown a priori, the min--max approach cannot be applied \citep{hussain2008financial, bhanja2018impact}. Instead, the prices are normalized with respect to their values at opening:
\begin{equation}\label{normalized_prices}
  \begin{aligned}
    \mathbf{V}_{t}^{\textrm{(Lo)}} &= \left[\frac{\mathop\textrm{Lo}(t-n-1)}{Op(t-n-1)},\cdots,\frac{\mathop\textrm{Lo}(t-1)}{\mathop\textrm{Op}(t-1)}\right]^{T},\\
    \mathbf{V}_{t}^{\textrm{(Cl)}} &= \left[\frac{\mathop\textrm{Cl}(t-n-1)}{\mathop\textrm{Op}(t-n-1)},\cdots,\frac{\mathop\textrm{Cl}(t-1)}{\mathop\textrm{Op}(t-1)}\right]^{T},\\ 
    \mathbf{V}_{t}^{\textrm{(Hi)}} &= \left[\frac{\mathop\textrm{Hi}(t-n-1)}{\mathop\textrm{Op}(t-n-1)},\cdots,\frac{\mathop\textrm{Hi}(t-1)}{\mathop\textrm{Op}(t-1)}\right]^{T}.
  \end{aligned}
\end{equation}
while the financial indicators are normalized with respect to their closing value on the previous day:
\begin{equation}\label{normalized_scaled_FI}
  \mathbf{V}_{t}^\textrm{(FI)} = \left[ \frac{\mathop\textrm{FI}(t-n)}{\mathop\textrm{FI}(t-n-1)},\cdots,\frac{\mathop\textrm{FI}(t)}{\mathop\textrm{FI}(t-1)}\right]^{T}.
\end{equation}
Feature extraction is performed on these normalized vectors with a restricted autoencoder.

\subsection{Restricted Stacked Autoencoder}\label{subsec:RSAE}
In order to implement machine learning algorithms efficiently, it is important to reduce the computational cost and time required for training \citep{meng2017relational}. For this purpose, feature extraction is performed to obtain highly informative abstract features with low dimensionality. The input feature vector contains eleven (11) features, namely: normalized low, high, and closing prices along with the eight normalized financial indicators, all taken at the same time $t$. These features are partially correlated and, as a result, carry redundant information. Feature extraction aims to remove such a redundancy while making the indicators more informative. The restricted, stacked autoencoder transforms recurrent events into low-dimension abstractions \citep{langr2019gans}.  

\begin{figure}[H]
  \centering
  \vspace{0.5cm}
  \includegraphics[width = 0.6\textwidth]{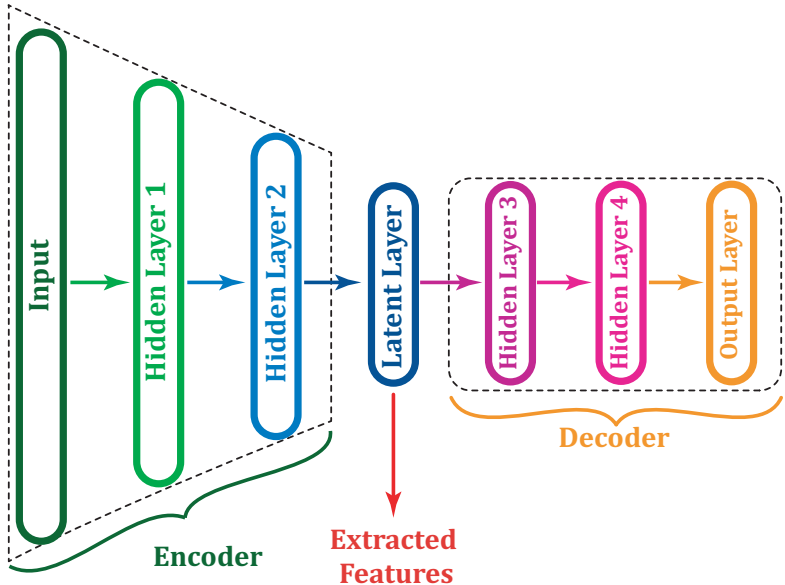}
  \caption{Restricted stacked autoencoder.}
  \label{fig:restricted_stacked_autoencoder}
\end{figure}

The restricted autoencoder consists of two parts, namely the encoder and the decoder. The encoder maps the input feature to a lower dimension via multiple stacked layers, collectively called the latent layer, while the decoder reconstructs low-dimensional features at the output. The decoder is designed to reconstruct only a subset of the original features; in our case, the low, close, and high price. In order to partially reconstruct the normalized input vector, each layer of the decoder has three (3) neurons which restrict the reconstruction to the normalized low, high, and closing prices. The network is under-complete, which means the latent layer has fewer nodes than the input layer \citep{bengio2015deep}. The latent layer creates a tensor of abstract features as $([V_{1}, V_{2}, V_{3}])$ . The structure of our RSAE is illustrated in Fig.~\ref{fig:restricted_stacked_autoencoder} 
In the next section, the correlation between financial assets is extracted using graph convolutional networks.

\section{Spectral Graph Convolutional Networks} \label{subsec:Spect_GNN}
Most of the data used in deep learning have a regular grid structure \citep{bronstein2017geometric}. One may mention, for instance, the regular pixel structure of an image or times series at fixed time intervals.  These data may easily be encoded for neural network input as vectors, matrices, or tensors. These structures are readily represented in Euclidean space, exhibit properties such as local-connectivity, shift-invariance, and compositionality, to mention just a few \citep{stone2017teaching}. As a result, the convolution operation is properly defined, thus providing the basis for convolutional neural networks \citep{henaff2015deep}.

In contrast, many entities, such as financial instruments, cannot be represented on a regular grid due to the complex nature of their interrelations. Indeed, financial instruments are correlated among themselves, and their structures may be better represented by a graph in which the nodes correspond to the financial instruments while the edges or links correspond to some pair-wise correlation function. It is worth noting that a single financial instrument, being a time series, embeds naturally in Euclidean space, but a plurality of financial instruments, because of their mutual correlations, resists Euclidean representation and instead is represented as a graph \citep{wu2019comprehensive, zhang2019graph}. Unfortunately, standard convolution cannot be directly extended to non-Euclidean geometries \citep{shuman2013emerging} therefore impeding its applicability to CNN. Nonetheless, as set out by the convolution theorem \citep{shuman2013emerging}, the convolution may be evaluated with the help of the Fourier transform: firstly, the Fourier transforms of both the input and the filter are evaluated; then, both transformations are multiplied element-by-element (the Hadamard product) and, finally, the inverse Fourier transform of the Hadamard product is taken. The convolution theorem remains valid under non-Euclidean geometry if the Fourier transform is properly defined \citep{shuman2013emerging, henaff2015deep}, thus allowing the application of CNN to non-Euclidean geometries.

The next section provides a mathematical formulation of spectral graph convolution.

\subsection{Graph Fourier transform} \label{sec:Graph_fourier}
Let us consider an undirected weighted graph $G = \{V, E, \mathbf{W}\}$ representing, for instance, $m$ financial instruments. This graph consists of $(|V| = m)$ vertices or nodes and $(|E| = n)$ edges or links. The edges represent the interrelations between the nodes while the weights are a pair-wise correlation function between nodes. This graph may be characterized by a weight adjacency matrix $\mathbf{W} \in \mathbb{R}^{m\times n}$:
\begin{equation}\label{eq:weight_matrix}
  \mathbf{W} = [w_{ij}]
\end{equation}
where $w_{ij}$ is the weight between node $i$ and $j$ \citep{shuman2016vertex}.
Such a graph is dynamic in the sense that the correlations vary at each time interval. The correlation is evaluated over a period consisting of $n$ time intervals. This period should be relatively short as the correlation tends to disappear rapidly due to randomization \citep{fama1965behavior, rocchi2017emerging}. For the correlation, a value of one indicates a perfect correlation while a value of zero indicates its absence. It is desirable that correlated nodes be topologically close to each other while uncorrelated nodes are more aloof. In order to reflect this behavior, the weights are defined as
\begin{equation}\label{eq:covariance_weight}
w_{ij} = 1-\mathop\textrm{corr}(V_{i},V_{j})_{[t-n+1,t]}
\end{equation}
where the correlation is evaluated on the interval $[t-n+1, t]$.
A graph may be represented by a Laplacian. The standard graph Laplacian is defined as
\begin{equation}\label{eq:Laplacian}
  \mathbf{L} = \mathbf{D} - \mathbf{W}
\end{equation}
where $\mathbf{D}\in \mathbb{R}^{n\times n}$ is diagonal degree matrix that indicates the degree of connectivity of each node $d_{i, j} = \sum w_{ij}$and , $\mathbf{W}$ is the weight matrix \citep{chung1997spectral}. This Laplacian is a discrete representation of the Laplacian operator which is encountered, for instance, in the heat and Schrödinger equations. As such, the Laplacian completely characterizes a Gaussian random walk on the graph \citep{felmer2012positive}. The Laplacian may also be defined as a symmetrical matrix \citep{kipf2016semi}:
\begin{equation}\label{eq:Norm_Laplacian}
	\mathbf{L}_{sym} = \mathbf{D}^{-\frac{1}{2}}(\mathbf{D}-\mathbf{W})\mathbf{D}^{-\frac{1}{2}}
\end{equation}
Symmetry is highly desirable as the eigenvalues ($0 = \lambda_{0} \leq \lambda_{1} \leq \cdots \leq \lambda_{n-1}$) associated with a symmetric matrix will then be real while the eigenvectors are mutually orthogonal, which greatly reduces the computational complexity. The eigendecomposition of a symmetrical Laplacian is given by
\begin{equation}\label{eq:norm_Lap_eig}
\mathbf{L}_{sym} = \mathbf{\Phi}\mathbf{\Lambda} \mathbf{\Phi}^{T}
\end{equation}
where $\mathbf{\Phi} = [\phi_{0}, \cdots, \phi_{n-1}]$ are the column-wise orthogonal eigenvectors:
\begin{equation}
\mathbf{\Phi}^{T}\mathbf{\Phi} = \mathbf{I}
\end{equation}
The eigenvalues are represented as a diagonal matrix $\mathbf{\Lambda} = \mathop\textrm{diag}(\lambda_{0}, \lambda_{1}, \cdots, \lambda_{\max}) \in \mathbb{R}^{n \times n}$. The eigenvectors corresponding to the smallest eigenvalues are the smoothest in the sense that $|\phi_{l}(i) - \phi_{l}(j)|$ is small for two neighboring nodes $i$ and $j$ \citep{shuman2011chebyshev}. As aforementioned, graph convolution may be defined using the Fourier transform and the convolution theorem \citep{shuman2013emerging}. The graph Fourier transform of a signal $\mathbf{x}$ on the graph is defined as 
\begin{equation}
	\mathcal{F}(\mathbf{x}) = \mathbf{\Phi}^{\textrm{T}}\mathbf{x} = \tilde{\mathbf{x}}
\end{equation}
where $\mathbf{\Phi}^{\textrm{T}}$ indicates the transpose of $\mathbf{\Phi}$; whereas the inverse graph Fourier transform is obtained by
\begin{equation}
	\mathcal{F}^{-1}(\tilde{\mathbf{x}}) = \mathbf{\Phi}\tilde{\mathbf{x}}) = \mathbf{x}
\end{equation}
where $\mathcal{F}$ denotes the Fourier transform, and $\tilde{\mathbf{x}})$ is the signal in the spectral domain. The latter results from the orthogonality of the eigenvectors. Therefore, the Fourier transform is achieved by projecting the signal onto the transpose of the eigenvector matrix while the inverse transform is obtained by projecting onto the eigenvectors per se. 
The graph convolution is introduced in the next section.
\subsection{Spectral Graph Convolution} \label{sec:spect_graph_conv}
Having defined the graph Fourier transform, one may immediately define the spectral graph convolution with the help of the convolution theorem. \citep{bronstein2017geometric}. The standard convolution is defined as
\begin{equation}\label{eq:standard_convolution}
  (\mathbf{f} \star \mathbf{g})(x) = \int_{\Omega} \mathbf{f}(x-x')\mathbf{g}(x')dx'
\end{equation}
According to the latter, the convolution of a signal $\mathbf{x}\in \mathbb{R}^{n}$ with a parametric filter ($g_{\theta}$) is defined as 
\begin{equation}\label{eq:convolution_Fourier}
  \mathbf{x} \star_{G} g_{\theta} = \mathcal{F}^{-1} [\mathcal{F}(\mathbf{x})\odot\mathcal{F}(g_{\theta})]
\end{equation}
Where $\theta \in \mathbb{R}^{n}$ is a parametrization of the filter and $\odot$ is the element-wise or Hadamard product. As stated in the previous section, the Fourier transform and its inverse may be expressed in terms of the eigenvectors of the Laplacian matrix. Therefore, the spectral graph convolution becomes:
\begin{equation}
	\mathbf{x} \star_{G} g_{\theta} = \mathbf{\Phi}(\mathbf{\Phi}^{\textrm{T}}\mathbf{x})\odot(\mathbf{\Phi}^{\textrm{T}}g_{\theta})
\end{equation}
Because of the orthogonality of the eigenvectors, this equation may be further simplified:
\begin{equation}\label{sec:graph Fourier}
\mathbf{y} = \mathbf{x} \star_{G} g_{\theta} = g_{\theta}(\mathbf{L})\mathbf{x} = \mathbf{\Phi}g_{\theta}(\mathbf{\Lambda}) \mathbf{\Phi}^{\textrm{T}}\mathbf{x} = \mathbf{\Phi} \underbrace{\mathop{\textrm{diag}}(g_{\theta})}_{\tilde{G}} \mathbf{\Phi}^{\textrm{T}}
\end{equation}
where the kernel is given by $g_{\theta}(\mathbf{\Lambda}) = \mathop\textrm{diag}(g_{\theta}(\mathbf{\lambda}))$.
The filters are parametrized in terms of a truncated power series over the eigenvalue matrix. Because this matrix is diagonal, only the diagonal element needs to be exponentiated, thus reducing the complexity of the calculation from $\mathcal{O}(N^{2})$, for a dense matrix, to $\mathcal{O}(N)$:
\begin{equation}
g_{\theta}(\mathbf{\Lambda}) = \sum_{k=0}^{K-1}\theta_{k}\mathbf{\Lambda}^{k}
\end{equation}
where $\theta_{k} \in \mathbb{R}^{K}$ are the filter’s parameters \citep{defferrard2016convolutional}. 
Most filters are not localized because their spectrum has an infinite expansion \citep{defferrard2016convolutional}. This is in contrast with the standard convolution for which the filters associated with CNN are always localized. Nonetheless, localization may be achieved if the filters are expressed in terms of Chebyshev polynomials of the first kind \citep{hammond2011wavelets, wu2019comprehensive}:

\begin{equation}
	g_{\theta}(\mathbf{\Lambda}) = \sum_{k=0}^{K-1}\theta_{k}T_{k}(\tilde{\mathbf{\Lambda}})
\end{equation}
where $\tilde{\mathbf{\Lambda}} = \frac{2\mathbf{\Lambda}}{\lambda_{max}}-\mathbf{I}_{n}$ is the rescaled eigenvalue matrix and $T_{k}(\tilde{\mathbf{\Lambda}})$ is the Chebyshev polynomial of order $k$. A truncated development of order $K-1$ corresponds to a filter that spans a $K$-ring neighborhood: from a 1-ring up to a $K$-ring. Each polynomial corresponds to a particular neighborhood: $T_{1}$ corresponds to a 1-ring neighborhood, $T_{2}$ corresponds to a 2-ring neighborhood while $T_{k}$ corresponds to a $k$-ring neighborhood away \citep{kipf2016semi, hammond2011wavelets}. The Chebyshev polynomial may be defined by a recurrence relation, thus reducing the computational complexity. Indeed, given:
\begin{equation}
\bar{\mathbf{x}}_{k} = T_{k}(\tilde{\mathbf{L}})\mathbf{x} \in \mathbb{R}^{n}
\end{equation}
where the scaled Laplacian is defined as
\begin{equation}
\tilde{\mathbf{L}} = \frac{2\mathbf{L}}{\lambda_{max}} - \mathbf{I}_{n}
\end{equation}
the recurrence relation is given by 
\begin{equation}
\bar{\mathbf{x}}_{k} = 2\tilde{\mathbf{L}}\bar{\mathbf{x}}_{k-1} - \bar{\mathbf{x}}_{k-2} 
\end{equation}

With $\bar{\mathbf{x}}_{0}= \mathbf{x}$ and $\bar{\mathbf{x}}_{1} = \tilde{\mathbf{L}}\mathbf{x}$.
As a result, the complexity associated with the filtering operation $\mathbf{y} = g_{\theta}(\mathbf{L})\mathbf{x} = [\bar{x}_{0}, \bar{x}_{1}, \cdots, \bar{x}_{K-1}]$ is $\mathcal{O}(K|E)$, where $|E|$ is the number of edges.

Our graph convolutional network is described in the next section.
\subsection{Graph Convolution Network} \label{sec:graph_conv_net}
The spectral graph convolution in the non-Euclidean domain is obtained by applying the graph Fourier transform and the convolution theorem to both the input signal and the convolving filter \citep{bruna2013spectral}, 
 \begin{figure}[H]
  \centering
  \vspace{0.5cm}
  \includegraphics[width = 0.6\textwidth]{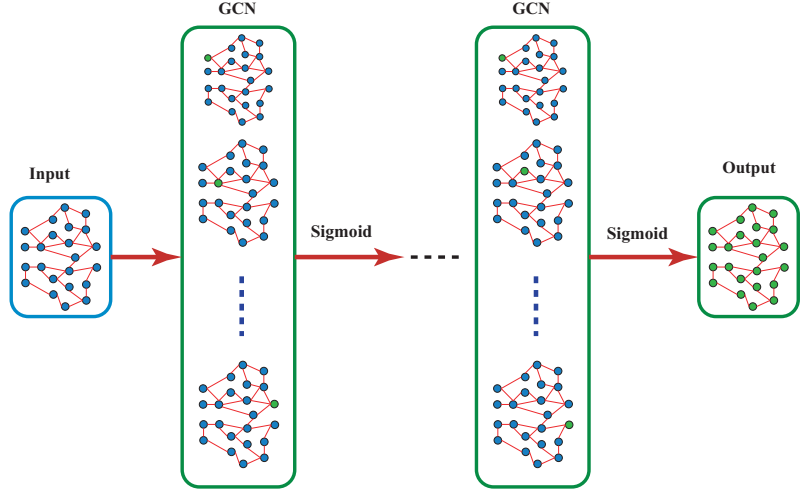}
  \caption{Graph Convolutional Neural Network}
  \label{fig:Graph_Convolutional_NN}
\end{figure}
The graph convolution extracts underlying local information by collecting the node information in a local neighborhood whose extension is determined by the order of the Chebyshev polynomial. In order to extract multi-scale substructure features, multiple graph convolution layers may be stacked as illustrated in Fig.~\ref{fig:Graph_Convolutional_NN}. The propagation rule for the multi-layer Graph Convolutional Network (GCN) is given by

\begin{equation}\label{eq:GCN}
  \mathbf{f}_{i}^{l+1} = \sigma\left(\sum_{j=1}^{p}\mathbf{\Phi}\mathbf{\hat{G}}_{i, j}\mathbf{\Phi}^\textrm{T}\mathbf{f}_{i}^{l}\right), \; i = 1,\cdots,q , \; j = 1, \cdots, p
\end{equation}

Where $\sigma$ denotes the nonlinear activation function ($l$) and $\hat{\mathbf{G}} = diag(g_{\theta}(\lambda))$. The number of features is indicated by ($q$) while the number of assets is given by ($p$).

\makebox[\linewidth]{%
\begin{minipage}{\dimexpr\linewidth-6em}
  \begin{algorithm}[H]
  \setcounter{AlgoLine}{0}
  \textbf{Input:} $\mathbf{x}_{t} = \left[v_{1}, v_{2}, v_{3}\right]^\textrm{T}$
  
  \textbf{Output:} Output feature vector $\leftarrow$ $\mathbf{F} = [\mathbf{F}_{1}, \mathbf{F}_{2}, \mathbf{F}_{3}]^\textrm{T} = [\mathbf{f}_{1}^{l+1}, \mathbf{f}_{2}^{l+1}, \mathbf{f}_{3}^{l+1}]^\textrm{T}$
  
  \KwData{$v_{1}, v_{2}, v_{3}:$ Extracted features from restricted stacked autoencoder for each asset}
  \KwData{$\mathbf{f}_{i}^{l+1}, \sigma:$ Transformed feature of layer $l+1$, Activation function [Sigmoid]}
  \textbf{Initialization:}
  \begin{enumerate}
    \item \textbf{Calculation of the graph Laplacian:}\\
      $\mathbf{L} = \mathbf{D}^{-1/2}(\mathbf{D}-\mathbf{W})\mathbf{D}^{-1/2}$
    \item \textbf{Eigendecomposition of the graph Laplacian:}\\
      $\mathbf{L} = \mathbf{\Phi}\mathbf{\Lambda}\mathbf{\Phi}^\textrm{T}$\\
      $\tilde{\mathbf{L}} = \frac{2\mathbf{L}}{\lambda_{max}} - \mathbf{I}_{n}$
    \item \textbf{Chebyshev polynomial (k, $\mathbf{x}_{t}$):}\\
      $\hat{\mathbf{x}}_{0} = \mathbf{x}_{t}$\\
      $\qquad$\eIf{$k == 1$}{
        $ \qquad \hat{\mathbf{x}}_{1} = \tilde{\mathbf{L}}\mathbf{x}$\;}{
        $ \qquad \hat{\mathbf{x}}_{k} = \tilde{\mathbf{L}}\hat{\mathbf{x}}_{k-1} - \hat{\mathbf{x}}_{k-2}$\;
       }
    \item \textbf{Kernel definition:}\\
      $\mathbf{\Phi}g_{\theta}(\mathbf{\Lambda})\mathbf{\Phi}^{T} = \mathbf{\Phi}\underbrace{\mathop\textrm{diag}(\hat{g}_{\theta})}_{\hat{\mathbf{G}}}\mathbf{\Phi}^{T}$
    \item \textbf{Graph Convolution:}\\
      $\mathbf{f}_{i}^{l+1} = \sigma\left(\sum_{j=1}^{p}\mathbf{\Phi}\mathbf{\hat{G}}_{i, j}\mathbf{\Phi}^{T}\mathbf{f}_{i}^{l}\right), \; i = 1,\cdots,q , \; j = 1, \cdots, p$\\
  \end{enumerate}
 \caption{Implementation of Graph Convolutional Network}
 \label{Alg:GCN}
\end{algorithm}
\end{minipage}}\\

 Algorithm~\ref{Alg:GCN} describes the graph convolutional network architecture. The output of the GCN is employed to train the actor--critic algorithm. The reinforcement learning approach is explained in detail in the following section.
\section{Deep Reinforcement Learning Framework}\label{sec:Deep_RL}
Deep reinforcement learning algorithms may be divided into three categories: critic-only, actor-only, and actor--critic methods. The critic-only approaches include, among others, Q-learning \citep{watkins1992q, bu2008comprehensive} and SARSA \citep{zhao2016deep}; the latter was employed in DeepBreath framework \citep{soleymani2020financial}. Critic-only methods rely on a state-action value function without explicitly defining a function over the policy space. Q-learning is an off-policy RL method in which the generated actions may be unrelated to the improvement of the policy. 
On the other hand, SARSA is an on-policy RL algorithm that seeks to improve the policy that generates the current actions \citep{andrew1999reinforcement}. On the other hand, actor-only methods optimize the cost over parameter space using a policy gradient approach. The actor-only methods converge much faster than critic-only methods \citep{grondman2012survey}. The actions generated by an actor-only method span a continuum, but the variance associated with the gradient is large, which results in a slower learning rate \citep{konda2000actor, sutton2000policy}. Finally, the actor--critic algorithm was developed in order to combine the advantages of both the actor-only and critic-only methods. Pseudo code for our actor--critic algorithm appears in section~\ref{sec:Actor_Critic}.
Reinforcement learning methods apply a sequence of actions taken by an agent, to ensure that the expected cumulative reward be optimum. In our framework, the actions correspond to the weight vector associated with the portfolio. 

\begin{equation}\label{eq:action}
  \mathbf{a}_{t} = \mathbf{w}_t.
\end{equation}

While the reward aims to maximize the expected return on investment. When considering the relation between the return rate Eq.~\ref{eq:ret_rate} and the transaction factor Eq.~\ref{eq:transaction_factor}, it may be concluded that the current action depends on the previous one. The state of the portfolio is obtained by concatenating the internal and external states: the former being the portfolio weight vector at time $t-1$ ($\mathbf{w}_{t-1}$) while the latter is the current feature tensor $X_{t}$:

\begin{equation}\label{eq:state}
  \mathbf{s}_t = [\mathbf{X}_t, \mathbf{w}_{t-1}]^\textrm{T}.
\end{equation}
The performance of each action is evaluated based on the achieved reward, aiming to increase the reward over a finite investment horizon $(\mathit{t_{f}+1})$. The reward is determined by the logarithmic accumulated return $\mathcal{R}$: Eq.~\ref{eq:log_return}, \ref{eq:ptf_final} and \ref{eq:transaction_factor}:

\begin{align}\label{eq:logarithmic_return}
  \mathcal{R}(\mathbf{s}_{1},\mathbf{a}_{1},\cdots ,\mathbf{s}_{t},\mathbf{a}_{t},\mathbf{s}_{t+1}) &= \frac{1}{t_{f}}\cdot \ln \left({\frac{P_{f}}{P_{0}}}\right) \nonumber\\
  &= \frac{1}{t_{f}} \cdot \sum^{t_{f}+1}_{t=1}\ln \left({\mu_{t}\cdot \mathbf{Y}_{t}\cdot \mathbf{w}_{t-1}}\right) = \frac{1}{t_{f}}\cdot \sum^{t_{f}+1}_{t=1} R_{t}.
\end{align}
The actor--critic algorithm used to learn the policy is described in the next section.
\subsection{Actor--Critic Architecture}\label{sec:Actor_Critic}
As the name suggests, the actor--critic algorithms--consists of two main components: the actor and the critic. The actor undertakes a sequence of actions following a policy while the critic evaluates the policy by assigning a performance index, called the value function, to each of the actor’s actions. The performance is evaluated using temporal differences (TD) \citep{sutton1988learning}. The critic approximates and updates the value function, which indicates the direction in which the actor’s policy parameters should be updated to improve performance. In this method, the policy update is derived directly from the value function, while the value function enforces the direction of the policy gradient at each time step \citep{grondman2012survey}. Actor--critic algorithms are capable of generating continuous actions while avoiding large policy gradient variance.

\begin{figure}[H]
  \centering
  \vspace{0.5cm}
  \includegraphics[width = 0.6\textwidth]{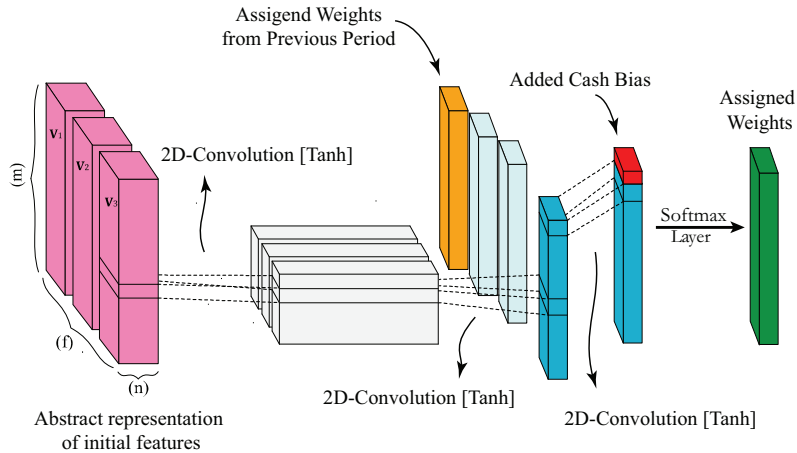}
  \caption{Architecture of actor}
  \label{fig:Actor_network}
\end{figure}

In this work, an actor--critic algorithm is employed in order to find the optimal policy for a higher return on investment. The actor undertakes the actions while the critic evaluates the state-value function as a success metric for the actions. The actor updates the policy in the direction suggested by the critic. The actor consists of a deep convolutional neural network, as illustrated in Fig.~\ref{fig:Actor_network}. Convolutional neural networks are known for their ability to learn complex patterns and efficiently implement policies \citep{jiang2017deep}. The output of the actor is the learned policy, which determines the portfolio allocation. The network is trained in two stages: initially, offline learning is performed based on a historical dataset using online stochastic batching \citep{jiang2017deep}, leading to the offline policy. Then, as new data become available, the policy is updated online. The architecture of the actor is shown in Algorithm~\ref{Alg:Actor}. The input tensor of this network consists of the high-level features learned from the restricted stacked autoencoder. The tensor consists of three (3) channels, with which three (3) matrices are associated. 

The number of rows in each matrix is determined by the number of assets ($m$), while the number of columns corresponds to the size of the trading window ($n$). The initial convolution is performed on the input tensor with a kernel of $1\times 3$ size, and then a $Tanh$ activation function is applied. The second convolutional layer has a kernel of size $1\times n$ aiming to obtain a vector of size $m$. The third layer is connected to the fourth layer through convolution with a kernel of $ 1\times 1$ size (obtaining one main channel) and a $Tanh$ activation function. The third layer consists of the previous actions (portfolio vector), i.e., the current action is computed based on both the current state and previous actions. The cash bias is added to the acquired action vector in the fourth layer, generating a vector of size $m+1$. The last layer normalizes the action vector using a Softmax function, resulting in a new portfolio weight distribution. All of the dimensions were determined by inspection.

\begin{figure}[H]
  \centering
  \vspace{0.5cm}
  \includegraphics[width = 0.6\textwidth]{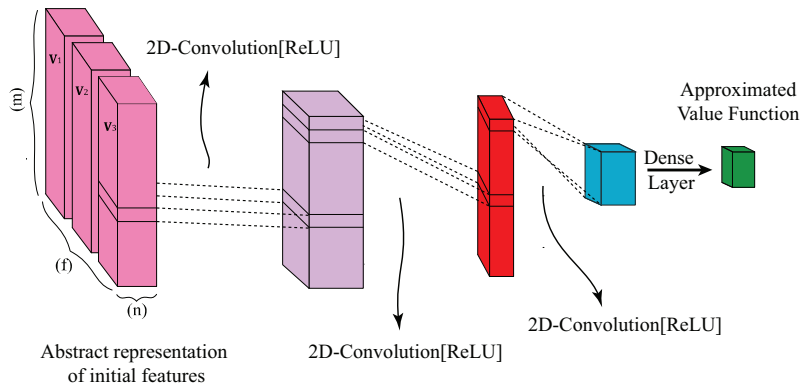}
  \caption{Critic architecture based on deep convolutional neural network}
  \label{fig:Critic_Network}
\end{figure}

Similarly, the architecture of the critic is described in Fig.~\ref{fig:Critic_Network}, and the corresponding pseudocode may be found in Algorithm~\ref{Alg:Critic}. The input consists of the current feature tensor $\mathbf{X}_{t}$. The size of the kernels for the first and second convolutional layers, is $1 \times 1$. This is followed by a $\mathop\textrm{ReLU}$ activation function. The third convolutional layer has a kernel of size $1 \times m$ and $\mathop\textrm{ReLU}$ activation function. Finally, the approximated value function is obtained by means of a densely connected layer. 
\makebox[\linewidth]{%
\begin{minipage}{\dimexpr\linewidth-6em}
  \begin{algorithm}[H]
  \textbf{Input:} $\mathbf{X}_{t} = \left[\mathbf{F}_{1}, \mathbf{F}_{2}, \mathbf{F}_{3}\right]^{\textrm{T}}$
  
  \textbf{Output:} Distributed Portfolio Weights $\leftarrow$ $\mathbf{W}_{ptf} = [\mathbf{w}_{C}, \mathbf{w}_{1}, \mathbf{w}_{2}, \cdots, \mathbf{w}_{m}]^\textrm{T}$

  \setcounter{AlgoLine}{0}
  \KwData{$f, m:$ Number of Features, Number of Assets}
  \textbf{Initialization:}\\
  \While{i $\leq$ n}{
  \begin{enumerate}
    \item \textbf{First Convolution:}\\
      Activation Function: Tanh \\
      Kernel size $\rightarrow$ (1,3)\\
      Kernel depth $\rightarrow$ $f$
    \item \textbf{Second Convolution:}\\
      Activation Function: Tanh \\
      Kernel size $\rightarrow$ (1,n)\\
      Kernel depth $\rightarrow$ $f$
    \item \textbf{Third Convolution:}\\
      Concatenate weights for previous and current time steps (Weight Matrix).\\
      $\mathbf{W} = [\mathbf{W}_{1}, \mathbf{W}_{2},\cdots, \mathbf{W}_{i}]^\textrm{T}$\\
      Activation Function: Tanh \\
      Kernel size $\rightarrow$ (1,1)\\
      Kernel depth $\rightarrow$ $f+1$
    \item \textbf{Adding Cash bias:}\\
      Concatenate Cash Bias and computed Weight Matrix
    \item \textbf{Softmax Layer:}\\
      Distributed Portfolio Weights
  \end{enumerate}}
 \caption{Implementation of Actor Algorithm}
 \label{Alg:Actor}
\end{algorithm}
\end{minipage}}\\

The actor is implemented with convolutional neural networks, including three convolutional layers and a final Softmax layer that generates the weights distribution vector. The current action at time $t$ is a function of the action at time $t-1$. The loss function of this network is evaluated based on the log-normal distribution of the policy $\pi_{\theta}$ and the temporal difference error (TD error) calculated by the critic network. The actor neural network follows a policy gradient approach, which updates the weight using a one-step TD error from Eq.~\ref{eq:TD_error}.

\makebox[\linewidth]{%
\begin{minipage}{\dimexpr\linewidth-6em}
  \begin{algorithm}[H]
  \textbf{Input:} $\mathbf{X}_{t} = \left[\mathbf{F}_{1}, \mathbf{F}_{2}, \mathbf{F}_{3}\right]^\textrm{T}$
  
  \textbf{Output:} Approximated value function $V_{v}$  
  
  \setcounter{AlgoLine}{0}
  \KwData{$f, m, n:$ Number of Features, Number of Assets, Trading Window Length}
  \textbf{Initialization:}\\
  \While{i $\leq$ n}{
  \begin{enumerate}
    \item \textbf{First Convolution:}\\
      Activation Function: ReLU \\
      Kernel size $\rightarrow$ (1,1)\\
      Kernel depth $\rightarrow$ $f$
    \item \textbf{Second Convolution:}\\
      Activation Function: ReLU \\
      Kernel size $\rightarrow$ (1,1)\\
      Kernel depth $\rightarrow$ $n$
    \item \textbf{Third Convolution:}\\
      Activation Function: ReLU \\
      Kernel size $\rightarrow$ (1,m)\\
      Kernel depth $\rightarrow$ $1$
    \item \textbf{Dense Layer}\\
      Value $(V_{v})$ Approximation
    \item \textbf{Training}\\
      Compute the TD error using Eq.~\ref{eq:TD_error}.
      
      \textit{Minimize the loss}\\
      Compute loss using Eq.~\ref{eq:temp_diff}
  \end{enumerate}}
 \caption{Implementation of Critic Algorithm}
 \label{Alg:Critic}
\end{algorithm}
\end{minipage}}\\

The critic consists of three convolutional neural networks.
The critic network is updated aiming to minimize, in Eq.~\ref{eq:temp_diff}, the mean square error (MSE) between the approximated value function $V_{v}$ and target value $V^{\pi}$.

\begin{equation}\label{eq:temp_diff}
  MSE = \|V^{\pi} - V_{v} \|^{2}
\end{equation}
The weights of the critic network are updated based on the gradient descent method, as shown in Algorithm~\ref{Alg:Critic}. The actor-critic framework is detailed in Algorithm~\ref{Alg:Actor_critic}. 

\begin{equation}\label{eq:TD_error}
  \delta_{v} = r(\mathbf{s}, \mathbf{a}) + \gamma V_{v}(\mathbf{s}') - V_{v}(\mathbf{s})
\end{equation}

where $r(\mathbf{s}, \mathbf{a})$ denotes immediate reward at state $\mathbf{s}$ resulting from action $\mathbf{a}$. The trade-off in between immediate and long term strategy is established by $\gamma$. The weights of the critic network are updated based on the gradient descent method as shown in Algorithm~\ref{Alg:Critic}. The actor-critic framework is detailed in Algorithm~\ref{Alg:Actor_critic}. 

\makebox[\linewidth]{%
\begin{minipage}{\dimexpr\linewidth-6em}
  \begin{algorithm}[H]
  \setcounter{AlgoLine}{0}
    \textbf{Input:} Initial network weights: $\theta_{a, init}, w_{c, init}$
    
    \KwData{$\alpha_{a}, \alpha_{c} \in [0,1]:$ Actor and critic learning rate}
    \KwData{$\theta_{a}, w_{c}:$ Actor and critic neural network weights}
    \KwData{$N_{t}:$ Number of iterations}
    \textbf{Initialization:}\\
    $iteration = 0$\\
    $\pi_{\theta}(\mathbf{s}, \mathbf{a}) \in \mathcal{A}$\\
    \While{iteration $\leq$ $N_{t}$}{
      $\Delta_{w} = \alpha_{c} \delta_{v} \nabla_{w} V(\mathbf{s})$\\
      $w_{c} \leftarrow w_{c} + \Delta_{w}$\\
      $\Delta_{\theta} = \alpha_{a} \delta_{v}\nabla_{\theta} \log \pi_{\theta}(\mathbf{s}, \mathbf{a})$ \\
      $\theta_{a} \leftarrow \theta_{a} + \Delta_{\theta}$\\
      \textit{\textbf{Update state and action:}}\\
      $\mathbf{s} \leftarrow \mathbf{s}'$\\
      $\mathbf{a} \leftarrow \mathbf{a}'$
    }
    \caption{Actor-Critic Algorithm}
    \label{Alg:Actor_critic}
  \end{algorithm}
\end{minipage}}\\
The actor--critic algorithm uses the Adam optimization method \citep{kingma2015adam} to update the weights for both the actor and the critic. The actor and the critic act alternately. At first, the weights associated with the networks are initialized randomly, and an initial admissible policy $\pi_{\theta}$ is calculated. The portfolio state is determined with respect to the initial action, and the value function is approximated. Then, the weights of the critic network are updated. Afterward, the weights of the actor network are updated using the TD error measured by the critic network. Finally, the state $(\mathbf{s})$ and action $(\mathbf{a})$ are updated. In other words, at each step, the actor network yields an action $(\mathbf{a})$ at state $(\mathbf{s})$, following the direction enforced by the critic network, in order to minimize the value function $(V_{v})$. 

\section{Online Learning}\label{sec:online_learning}

The proposed framework is trained using historical data (prior to current calendar time) and the current flow of data pertaining to offline and online learning, respectively. The offline learning is performed using online stochastic batching, described in \citep{soleymani2020financial}. Financial data in the stock market may be assimilated into time series. As a result, unanticipated changes in their underlying distribution may occur, a phenomenon known as concept drift \citep{gama2014survey}. These may be generated by exogenous factors such as panic reactions (e.g. the Covid-19 pandemic) \citep{ashraf2020stock, zhang2020financial}, and political disputes (e.g. US--China trade war) \citep{cavalcante2015approach}. Such factors may profoundly affect the actions taken by investors. In this work, concept drift is addressed using a passive detection approach where the agent is first trained based on historical data, and then the learned policy is continuously updated as new data arrive, thus allowing for concept drift without discarding useful knowledge \citep{soleymani2020financial}. 

\begin{figure}[H]
  \centering
  \vspace{0.5cm}
  \includegraphics[width = 0.6\textwidth]{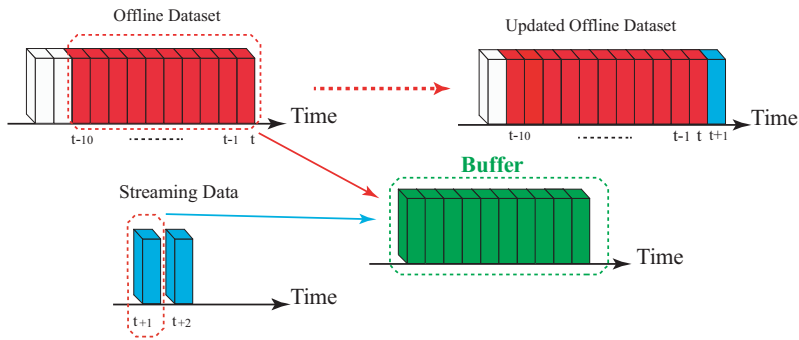}
  \caption{Structure of the buffer that stores data from the offline dataset as well as the flow of online data for online learning}
  \label{fig:buffer}
\end{figure}

The motivation of the present approach is twofold: extreme events are very often recurrent, which means that a solution to a current financial crisis may be brought from past crises \citep{hu2015concept}. Secondly, the passive detection approach discards historical patterns that are not repeated after a certain amount of time \citep{gama2004learning}, thus preventing any unstable, short-term investment strategies based on outliers. All of the neural networks in our portfolio management framework (RSAE, GCN, actor--critic), are first trained offline on historical data. Then, the weights are updated online as new data become available. The online learning is managed by a buffer, as shown in Fig.~\ref{fig:buffer}. The buffer stores the financial data from the last ten days, along with the current business day. Then the stored set containing data from eleven days is used to train and update the networks. At the end of the business transaction (current day), the current financial data are added to the offline dataset.

\section{Experimental Results}\label{sec:EXPresults}
The assets forming our portfolio appear in Table~\ref{tab:financial_assets}. This dataset consists of various financial time series which cover a period ranging from January 2, 2002, up to March 24, 2020; details about each time series may be found in Table~\ref{tab:data_range}. In reinforcement learning, actions are taken to minimize the risk while increasing the return on investment. Therefore, to further investigate the risk involved with the investment policy, the performance of the proposed framework is evaluated using three of the most celebrated risk evaluation metrics \citep{bessler2017multi}: namely maximum drawdown (MDD), Sharpe ratio $(S_{r})$,
and Conditional-Value-at-Risk (CVaR) \citep{rockafellar2000optimization}.
The maximum drawdown metric was first proposed by \citep{magdon2004maximum}. This metric is defined as the difference in between the maximum peak value $P$ and the minimum value $L$ reached by the portfolio for a certain period of time before the next peak occurs. Technically, this indicator measures the amplitude of a loss over a period of time. A low maximum drawdown indicates a small loss. 
\begin{equation}\label{eq:MDD}
  MDD = \frac{P-L}{P}.
\end{equation}
The Sharpe ratio was introduced by \citep{sharpe1994sharpe}. It measures the volatility of investment as a ratio between the expected return on investment and the risk:
\begin{equation}\label{eq:Sharpe_Ratio}
  S_{r} = \frac{\mathnormal{E}\left[R_{f}-R_{b}\right]}{\sigma_{d}}.
\end{equation}
where $\sigma_{d}$ is the standard deviation associated with the asset excess return (or volatility) where $(E[R_{f}–R_{a}])$ is the expected differential return. A high return on investment is achieved when the Shape ratio is greater than one. In other words, this metric helps investors to better identify the returns associated with high-risk actions.

In order to further analyze the risk associated with portfolio optimization, risk assessment criteria called Value-at-Risk (VaR) and Conditional-Value-at-Risk (CVaR) are adopted here. The former measures and control the risk with respect to percentiles of the loss distribution, while later estimates the tail risk of a portfolio \mbox{\citep{alexander2006minimizing}}. Nonetheless, the use of VaR is limited due to lack of subadditivity and being non-convex and non-smooth with multiple local minima \mbox{\citep{acerbi2002spectral}}. On the other hand, the CVaR renders a convex optimization problem \mbox{\citep{cornuejols2006optimization}}.
\begin{equation}\label{eq:CVAR}
  CVaR_{\alpha}(X) = \mathop{\mathbb{E}}[X|X\geq VaR_{\alpha}(X)]
\end{equation}

The historical training set consists of four different time series as shown in Table.~\ref{tab:data_range}. Each training set spans a different period, allowing our system to learn from various historical contexts. Each training batch covers 90 consecutive days. Initially, the portfolio consists only of cash, so the portfolio composition is determined entirely by the agent from the start. As mentioned earlier, the assets forming the portfolio appear in Table.~\ref{tab:financial_assets}. These assets were selected in order to reflect various segments of the economy, such as the pharmaceuticals industry, financial services, manufacturing, healthcare, energy, and consumer services. This portfolio allows us to further investigate the interrelation among financial instruments that may be perceived as statistically independent at first sight. 
\begin{table}[H]
\small
\centering
\caption{Composition of our Portfolio for the period in between January 2, 2002, and February 10, 2020.}
\label{tab:financial_assets}
\begin{tabular}{cl}
\toprule
\textbf{Sector} & \textbf{Name} \\ \midrule
\multirow{8}{*}{\textit{\textbf{Technology}}} & Apple.Inc (AAPL) \\ 
 & Cisco Systems Inc. (CSCO) \\
 & Intel Corporation (INTC) \\ 
 & Oracle Corporation (ORCL) \\ 
 & Microsoft Corporation (MSFT) \\ 
 & International Business Machines Corporation (IBM) \\
 & Honeywell International Inc. (HON) \\ 
 & Verizon Communications Inc. (VZ) \\ 
 \midrule
\multirow{4}{*}{\textit{\textbf{Financial Services}}} & Manulife Financial Corporation (MFC) \\
 & JPMorgan Chase \& Co. (JPM) \\ 
 & Bank of America Corporation (BAC) \\ 
 & The Toronto-Dominion Bank (TD) \\ 
\midrule
\multirow{4}{*}{\textit{\textbf{Industries}}} & 3M Company (MMM) \\ 
 & Caterpillar Inc. (CAT) \\ 
 & The Boeing Company (BA) \\ 
 & General Electric Company (GE) \\ 
\midrule
\multirow{2}{*}{\textit{\textbf{Consumer Defensive}}} & Walmart Inc. (WMT) \\
 & The Coca-Cola Company (KO) \\
\midrule
\multirow{2}{*}{\textit{\textbf{Consumer Cyclical}}} & The Home Depot, Inc. (HD) \\ 
 & Amazon.com, Inc. (AMZN) \\ 
\midrule
\multirow{4}{*}{\textit{\textbf{Healthcare}}} & Johnson \& Johnson (JNJ) \\ 
 & Merck \& Co., Inc. (MRK) \\ 
 & Pfizer Inc. (PFE) \\ 
 & Gilead Sciences, Inc. (GILD) \\
\midrule
\multirow{4}{*}{\textit{\textbf{Energy}}} & Enbridge Inc. (ENB) \\
 & Chevron Corporation (CVX) \\ 
 & BP p.l.c. (BP) \\ 
 & Royal Dutch Shell plc (RDSB.L) \\ 
\bottomrule
\end{tabular}
\end{table}

\begin{table}[H]
\small
\centering
\caption{Training and test sets}
\label{tab:data_range}
\begin{tabular}{ccc}
\toprule
\textbf{ID} & \textbf{Training Set} & \textbf{Test Set} \\
\midrule
\textit{\textbf{Test 1}} & 2002-01-02 to 2009-04-16 & 2010-03-15 to 2010-07-21 \\ 
\textit{\textbf{Test 2}} & 2002-01-02 to 2012-12-06 & 2013-11-04 to 2014-03-14 \\ 
\textit{\textbf{Test 3}} & 2002-01-02 to 2016-08-01 & 2017-06-28 to 2017-11-02 \\ 
\textit{\textbf{Test 4}} & 2002-01-02 to 2018-10-19 & 2019-06-09 to 2019-10-16 \\ 
\textit{\textbf{Test 5}} & 2002-01-02 to 2019-06-23 & 2019-11-12 to 2020-03-24 \\
\bottomrule
\end{tabular}
\end{table}

\begin{figure}[H]
  \centering
  \includegraphics[width = 0.6\textwidth]{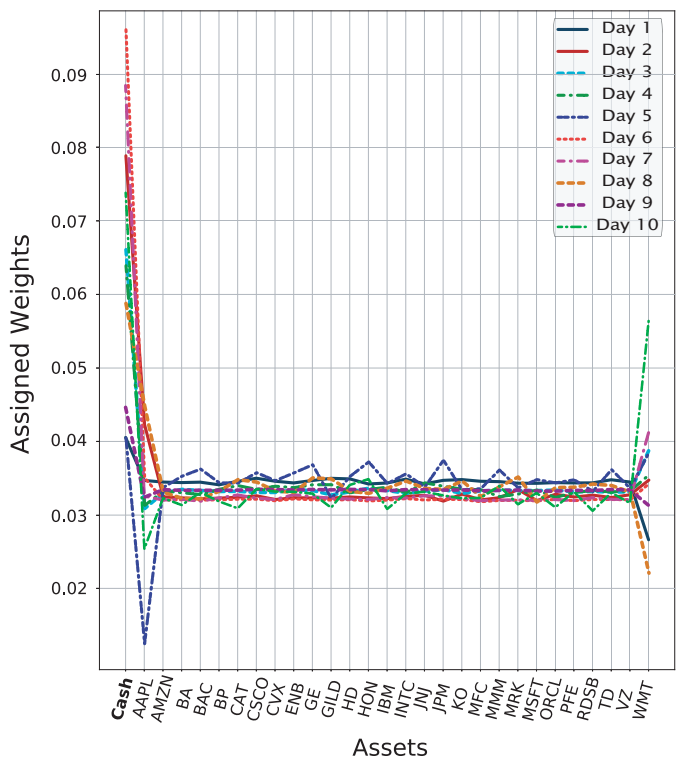}
  \caption{Evolution of the portfolio weights during the first 10 days of the online learning process.}
  \label{fig:online_learning}
\end{figure}
Our online learning results are shown in Fig.~\ref{fig:online_learning}, where the pre-trained model from the last training set is employed, and the weights for the online model are updated as new data becomes available (for each trading day). The learning is conducted over 10 trading days from 2020-03-24 to 2020-04-06. Fig.~\ref{fig:online_learning} illustrates the weight fluctuations during the online learning process. Initially, the lowest weight corresponds to WMT, while the rest of the assets are uniformly distributed. Then, as the process unfolds, the weights associated with assets such as APPL and WMT are subject to great variations, while others, such as ORCL, PFE, and RDSB, are not. The most stable assets, in terms of their weights in the portfolio, are employed by the agent to achieve a stable return on investment by hedging the risk. More volatile assets, in terms of their weights, are exploited by the agent for their leverage effect \citep{bouchaud2001leverage}. 

\begin{figure}[H]
  \centering
  \includegraphics[width = 0.75\textwidth]{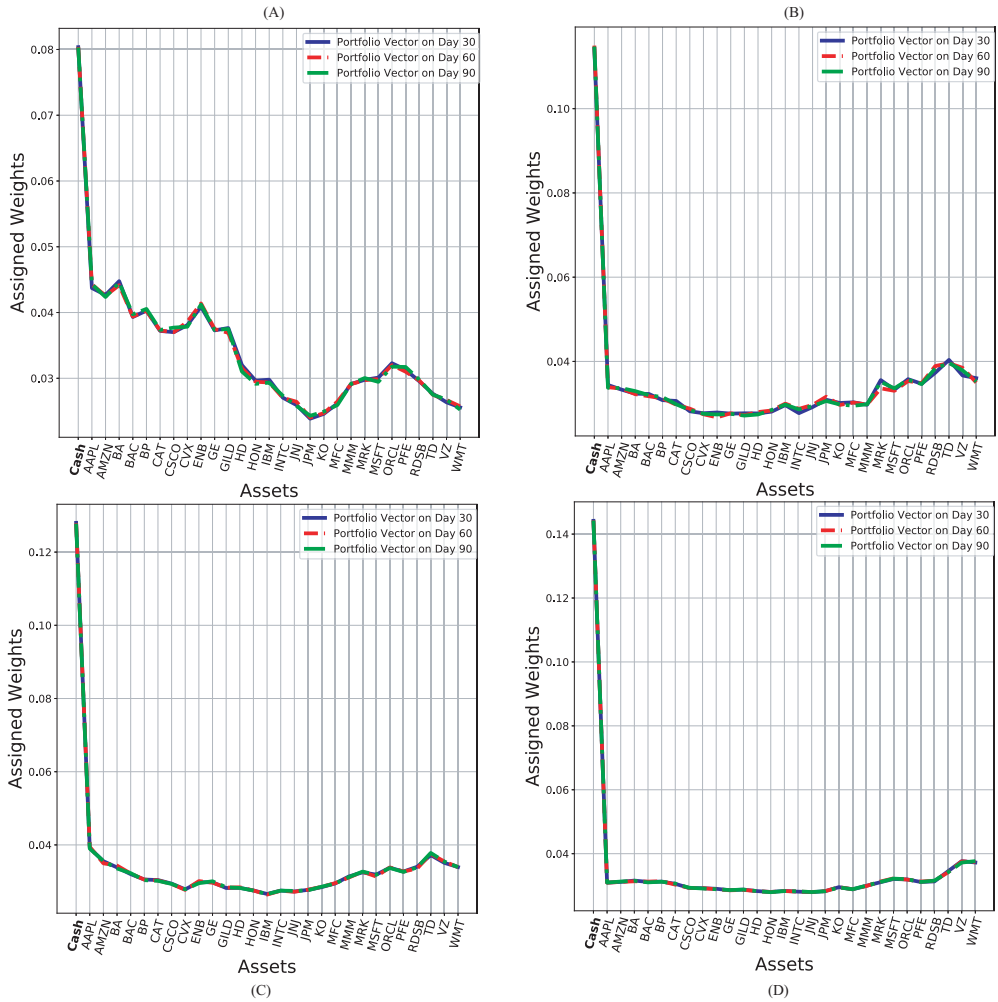}
  \caption{Portfolio weight distributions at the end of the investment horizon for four test sets.}
  \label{fig:test_weights}
\end{figure}

\begin{figure}[H]
  \centering
  \includegraphics[width = 0.65\textwidth]{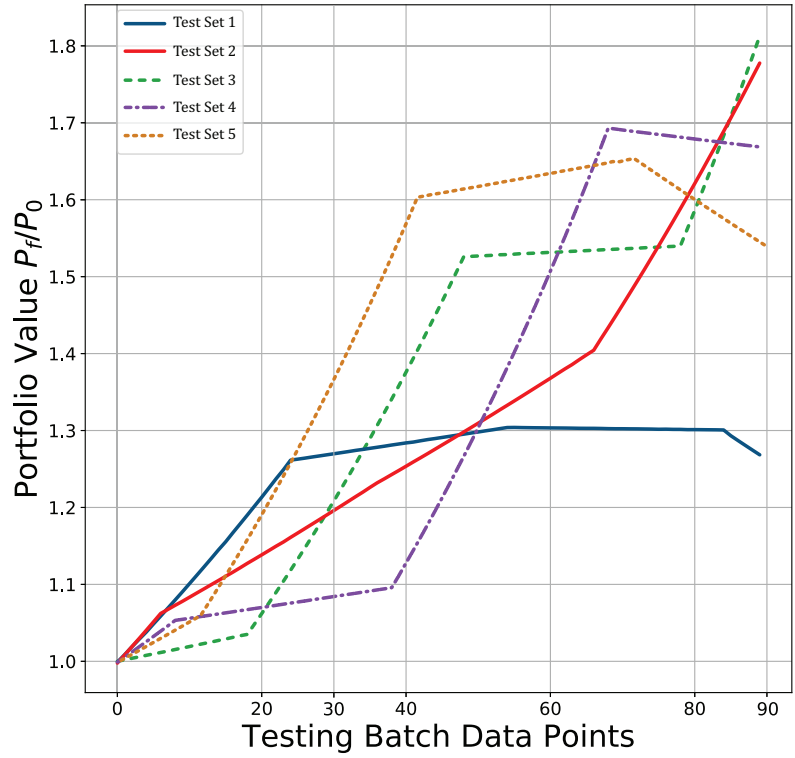}
  \caption{Relative portfolio values after respectively 30, 60, and 90 days of trading for all five test sets.}
  \label{fig:test_ptf}
\end{figure}

Fig.~\ref{fig:test_weights} illustrates the weight distribution for the four test sets, as defined in Table~\ref{tab:data_range}, after respectively 30, 60, and 90 days of investment. One may notice that the weights are more equally distributed for the last testing set. The behavior may be explained by the fact that the agent aims to mitigate risk by distributing the funds among the various assets. The resulting portfolio values are reported in Fig.~\ref{fig:test_ptf}. The portfolio value for the first test set is highlighted in blue. During the first 23 days, the portfolio value increases by $23.8\%$ and then, after 60 days, climbs gradually to $30.33\%$. Then, the portfolio value remains stable while, during the last five days, the growth decreases to $26.23\%$ after 90 days of trading. The evolution of the portfolio for the second test set consistently increases, showing a growth of $79.57\%$ after 90 days. As for the third test, the growth reaches $83.97\%$ after 90 days, the highest among all test sets. Finally, the fourth test set achieves $68.8\%$ growth after being 68 days on the market before gradually dropping to $60.81\%$ after 90 trading days. From Fig.~\ref{fig:test_ptf}, one may conclude that the proposed framework is particularly suitable for medium-term investments. 

\begin{figure}[H]
  \centering
  \includegraphics[width = 0.55\textwidth]{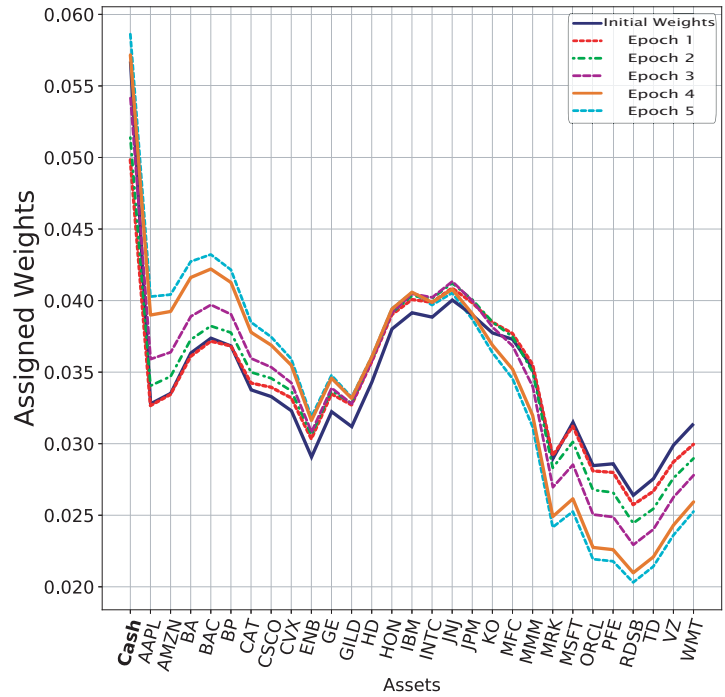}
  \caption{Final portfolio weights distributions for the fifth training set.}
  \label{fig:train_covid_90}
\end{figure}

\begin{figure}[H]
  \centering
  \includegraphics[width = 0.6\textwidth]{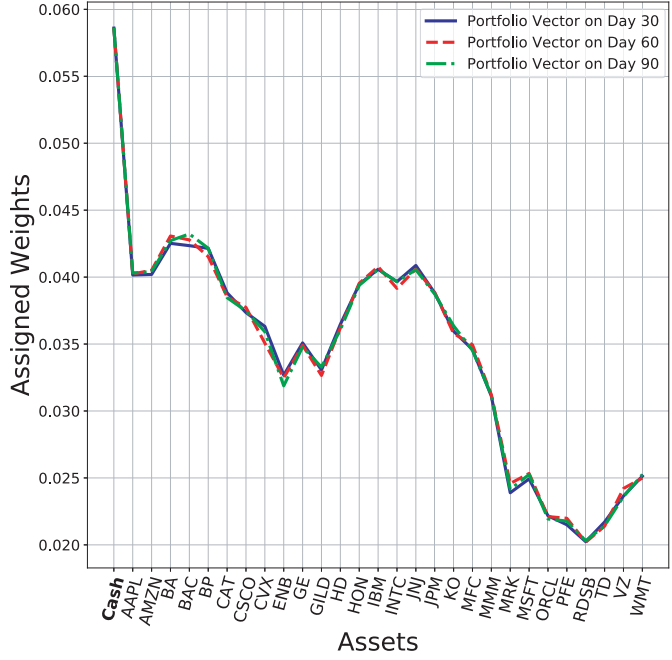}
  \caption{Final portfolio weight distributions for the fifth test set.}
  \label{fig:test_covid_90}
\end{figure}
\begin{table}[H]
\small
\centering
\caption{Maximum drawdown (MDD) and Sharpe ratio ($S_{r}$) for DeepPocket, the Dow Jones Industrial (DJI), Euro Stoxx 50, Nasdaq, and S\&P500 for all five test sets.}
\label{tab:return_sharpe}
\begin{tabular}{cccccccc}
\toprule
\textbf{ID} & \textbf{\begin{tabular}[c]{@{}c@{}}Investment \\ Duration\end{tabular}} & \textbf{\begin{tabular}[c]{@{}c@{}}MDD (\%)\\ Algorithm\end{tabular}} & \textbf{\begin{tabular}[c]{@{}c@{}}$S_{r}$\\ Algorithm\end{tabular}} & \textbf{\begin{tabular}[c]{@{}c@{}}$S_{r}$\\ DJI\end{tabular}} & \textbf{\begin{tabular}[c]{@{}c@{}}$S_{r}$\\ FEZ\end{tabular}} & \textbf{\begin{tabular}[c]{@{}c@{}}$S_{r}$\\ Nasdaq\end{tabular}} & \textbf{\begin{tabular}[c]{@{}c@{}}$S_{r}$\\ S\&P500\end{tabular}} \\ \midrule
 & 30 Days & 21.16 & {\color[HTML]{333333} 2.89} & 2.101 & -3.80 & 1.766 & 1.66 \\
 & 60 Days & 23.31 & 3.11 & -1.44 & -2.133 & -1.62 & -1.55 \\
\multirow{-3}{*}{\textit{\textbf{Test Set 1}}} & 90 Days & 23.31 & 2.95 & -1.458 & -1.722 & -1.362 & -1.42 \\ \midrule
 & 30 Days & 16.096 & {\color[HTML]{333333} 3.41} & 2.752 & -1.151 & 1.746 & 1.66 \\
 & 60 Days & 26.64 & 3.507 & -1.518 & -2.814 & 0.222 & -1.65 \\
\multirow{-3}{*}{\textit{\textbf{Test Set 2}}} & 90 Days & 43.74 & 3.84 & 0.949 & -0.418 & 2.261 & 1.83 \\ \midrule
 & 30 Days & 16.18 & 3.18 & 0.755 & -1.94 & -0.3914 & -1.070 \\
 & 60 Days & 34.68 & 2.53 & 2.23 & -0.987 & 0.755 & 1.27 \\
\multirow{-3}{*}{\textit{\textbf{Test Set 3}}} & 90 Days & 44.83 & 3.39 & 3.018 & 1.65 & 2.819 & 2.43 \\ \midrule
\cellcolor[HTML]{FFFFFF} & \cellcolor[HTML]{FFFFFF}30 Days & \cellcolor[HTML]{FFFFFF}7.71 & \cellcolor[HTML]{FFFFFF}3.24 & 1.75 & 1.93 & 1.951 & 2.031 \\
\cellcolor[HTML]{FFFFFF} & \cellcolor[HTML]{FFFFFF}60 Days & \cellcolor[HTML]{FFFFFF}32.70 & \cellcolor[HTML]{FFFFFF}3.67 & -0.520 & -2.518 & -0.0521 & -0.285 \\
\multirow{-3}{*}{\cellcolor[HTML]{FFFFFF}\textit{\textbf{Test Set 4}}} & \cellcolor[HTML]{FFFFFF}90 Days & \cellcolor[HTML]{FFFFFF}40.93 & \cellcolor[HTML]{FFFFFF}2.47 & 1.6134 & 0.114 & 0.981 & 1.42 \\ \midrule
\rowcolor[HTML]{FFFFFF} 
\multicolumn{1}{l}{\cellcolor[HTML]{FFFFFF}} & 30 Days & 26.34 & 3.107 & 2.51 & 1.65 & 3.300 & 3.014 \\
\rowcolor[HTML]{FFFFFF} 
\multicolumn{1}{l}{\cellcolor[HTML]{FFFFFF}} & 60 Days & 38.794 & 2.64 & 2.59 & 0.757 & 3.265 & 3.03 \\
\rowcolor[HTML]{FFFFFF} 
\multicolumn{1}{l}{\multirow{-3}{*}{\cellcolor[HTML]{FFFFFF}\textit{\textbf{Test set 5}}}} & 90 Days & 39.53 & 2.301 & -3.78 & -3.678 & -2.314 & -3.21 \\ \bottomrule
\end{tabular}
\end{table}

\begin{figure}[H]
  \centering
  \includegraphics[width = 0.6\textwidth]{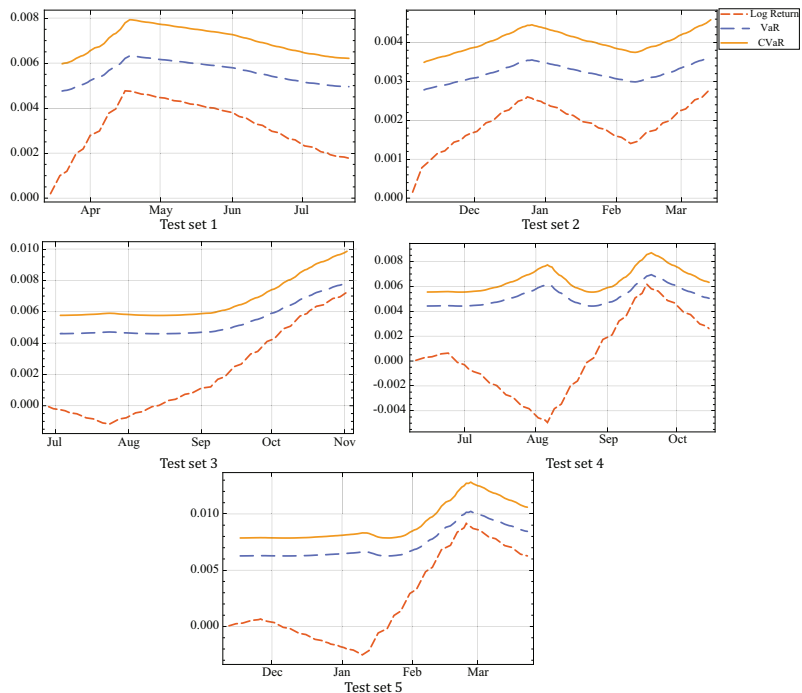}
  \caption{Risk assessment using CVaR and VaR criteria for all five test sets}
  \label{fig:CVaR_Norm}
\end{figure}

Our results are summarized in Table~\ref{tab:return_sharpe}. With the exception of Test Set 1, the Sharpe ratio is either very good or excellent for all four remaining test sets. In general, it is recognized that a Sharpe ratio below one is sub-optimal; a ratio between one and two is fair, and a ratio between two and three is considered very good. Any ratio over three is considered excellent \citep{Sharpe_Ratio_Evaluation}. In general, online training over a larger time window results in a more efficient policy in terms of return on investment. Our frameworks seems particularly adapted for medium-term investments as the value of the portfolio usually peaks after 80 to 90 trading days. 

\begin{figure}[H]
  \centering
  \includegraphics[width = 0.6\textwidth]{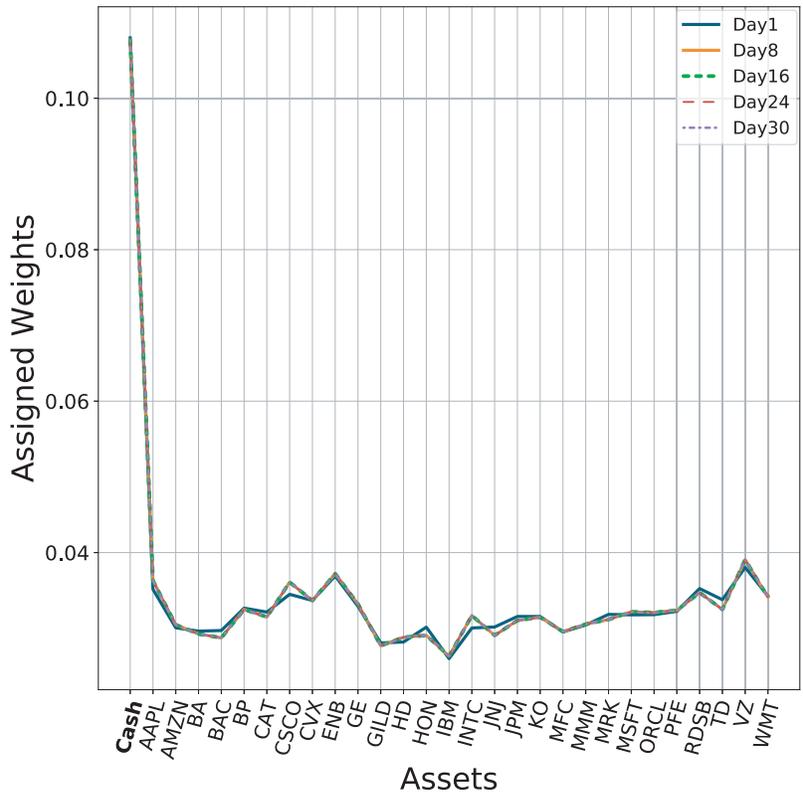}
  \caption{Evolution of the portfolio weights over a period of 30 days following the Covid-19 crisis.}
  \label{fig:test_weight_covid_30}
\end{figure}
\begin{table}[H]
\small
\centering
\caption{Portfolio weights distribution over a period of 30 days following the Covid-19 crisis(from Fig.~\ref{fig:test_weight_covid_30})}
\label{tab:30days_ptf_new_covid}
\begin{tabular}{cccccc}
\toprule
\textbf{ID} & \textbf{Day 1} & \textbf{Day 8} & \textbf{Day 16} & \textbf{Day 24} & \textbf{Day 30} \\ \midrule
\textbf{Cash} & 0.10905 & 0.10878 & 0.10845 & 0.10807 & 0.10735 \\ \midrule
\textbf{AAPL} & 0.0349 & 0.0361 & 0.0362 & 0.0365 & 0.037 \\ \midrule
\textbf{AMZN} & 0.0299 & 0.0305 & 0.03052 & 0.03055 & 0.03055 \\ \midrule
\textbf{BA} & 0.0296 & 0.0292 & 0.02918 & 0.02918 & 0.02918 \\ \midrule
\textbf{BAC} & 0.0297 & 0.0286 & 0.0287 & 0.0288 & 0.02885 \\ \midrule
\textbf{BP} & 0.0327 & 0.0322 & 0.0324 & 0.0322 & 0.0322 \\ \midrule
\textbf{CAT} & 0.0322 & 0.03164 & 0.0315 & 0.03185 & 0.03185 \\ \midrule
\textbf{CSCO} & 0.0344 & 0.03616 & 0.03605 & 0.03602 & 0.03602 \\ \midrule
\textbf{CVX} & 0.03368 & 0.03367 & 0.03367 & 0.03367 & 0.03367 \\ \midrule
\textbf{ENB} & 0.0367 & 0.037 & 0.037 & 0.03724 & 0.03724 \\ \midrule
\textbf{GE} & 0.0328 & 0.033 & 0.0328 & 0.0326 & 0.0326 \\ \midrule
\textbf{GILD} & 0.0283 & 0.0279 & 0.0278 & 0.0279 & 0.02792 \\ \midrule
\textbf{HD} & 0.0281 & 0.0288 & 0.0291 & 0.0292 & 0.0292 \\ \midrule
\textbf{HON} & 0.0301 & 0.0291 & 0.029 & 0.0289 & 0.0289 \\ \midrule
\textbf{IBM} & 0.0259 & 0.0261 & 0.0262 & 0.0266 & 0.0266 \\ \midrule
\textbf{INTC} & 0.0299 & 0.0317 & 0.0316 & 0.03158 & 0.03158 \\ \midrule
\textbf{JNJ} & 0.0302 & 0.02906 & 0.02908 & 0.02915 & 0.02915 \\ \midrule
\textbf{JPM} & 0.0316 & 0.03097 & 0.031 & 0.0312 & 0.0312 \\ \midrule
\textbf{KO} & 0.0315 & 0.03148 & 0.03143 & 0.03141 & 0.03141 \\ \midrule
\textbf{MFC} & 0.0297 & 0.0297 & 0.0297 & 0.0295 & 0.0295 \\ \midrule
\textbf{MMM} & 0.0303 & 0.03046 & 0.03058 & 0.03062 & 0.03062 \\ \midrule
\textbf{MRK} & 0.0318 & 0.03125 & 0.0313 & 0.0313 & 0.0313 \\ \midrule
\textbf{MSFT} & 0.03172 & 0.0321 & 0.03211 & 0.0322 & 0.0322 \\ \midrule
\textbf{ORCL} & 0.0321 & 0.03208 & 0.03208 & 0.03207 & 0.03207 \\ \midrule
\textbf{PFE} & 0.03225 & 0.03235 & 0.0325 & 0.0321 & 0.0321 \\ \midrule
\textbf{RDSB} & 0.0351 & 0.0348 & 0.03475 & 0.03471 & 0.03471 \\ \midrule
\textbf{TD} & 0.0338 & 0.0324 & 0.0325 & 0.0326 & 0.03275 \\ \midrule
\textbf{VZ} & 0.0378 & 0.039 & 0.0389 & 0.0389 & 0.0389 \\ \midrule
\textbf{WMT} & 0.0342 & 0.0339 & 0.0339 & 0.03338 & 0.03338 \\ \bottomrule
\end{tabular}
\end{table}

\begin{figure}[H]
  \centering
  \includegraphics[width = 0.6\textwidth]{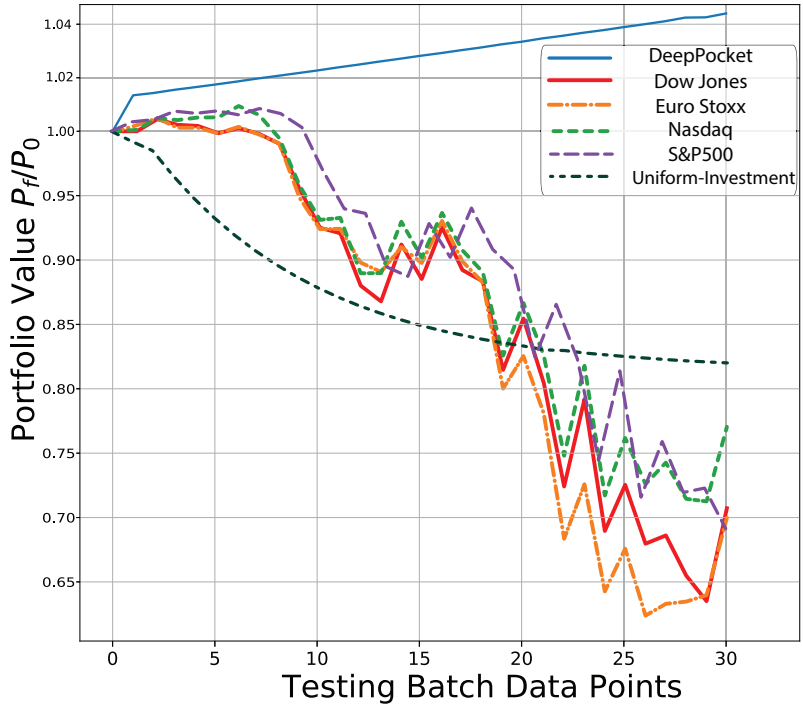}
  \caption{Evolution of the portfolio value over a period of 30 days following the Covid-19 crisis.}
  \label{fig:30days_ptf_new_covid}
\end{figure}
In order to test the robustness of our framework, DeepPocket managed the portfolio in the depths of the coronavirus crisis (Covid-19). The investments were performed from 11 February 2020, 7 days before the crisis, until 23 March 2020, while the stock market was witnessing record falls. The weight distribution during this time period is shown in Fig.~\ref{fig:test_weight_covid_30} and Table.~\ref{tab:30days_ptf_new_covid}. The evolution of the portfolio is illustrated in Fig.~\ref{fig:30days_ptf_new_covid}. The objective of this test was to evaluate the performance of the agent in unprecedented circumstances as the stock market was entering a bear market \citep{barsky1990bull}. As shown in Fig.~\ref{fig:30days_ptf_new_covid}, despite the debacle in worldwide financial markets, DeepPocket managed to increase the value of the portfolio by $4.46\%$. This is even more remarkable when considering that the agent was trained online on a bull market during a sharp rise. This test shows the agent’s adaptation capacity under the worst possible conditions. During this period, some assets were more affected than others: indeed, while IBM, GILD, JPM, and HON were in sharp decline, AMZN and WMT were rapidly recovering.

In comparison, as illustrated by Fig.~\ref{fig:30days_ptf_new_covid}, had one invested evenly across these assets, the portfolio value would have shrunk by $17.83\%$. Moreover, during the same period, the Dow Jones Industrial (DJI) lost $29.36\%$ of its value. Therefore, DeepPocket clearly outperforms uniform-investment and benchmark indexes in times of crisis. Our framework has the ability to remember remote events because of, among other reasons, gradual concept drift \citep{gama2014survey}. The fact that DeepPocket could manage the coronavirus crisis is to be found, in part, in such a long-term memory mechanism. Indeed, the offline training, which was performed on historical data, involved the early 2000's recession where the market reached a low point in 2002 as well as the 2009 market crisis \citep{langdon2002us, farmer2012stock}. The knowledge gained by DeepPocket over these two crisis periods was instrumental in the management of the coronavirus crisis.

\begin{table}[H]
\small
\centering
\caption{Return on investment for DeepPocket, the Dow Jones Industrial, Euro Stoxx 50, Nasdaq, and S\&P500 for all five test sets.}
\label{tab:RoI}
\begin{tabular}{ccccccc}
\toprule
\textbf{ID} & \textbf{\begin{tabular}[c]{@{}c@{}}Investment \\ Duration\end{tabular}} & \textbf{\begin{tabular}[c]{@{}c@{}}ROI(\%)\\ DeepPocket\end{tabular}} & \textbf{\begin{tabular}[c]{@{}c@{}}ROI(\%)\\ DJI\end{tabular}} & \textbf{\begin{tabular}[c]{@{}c@{}}ROI(\%)\\ FEZ\end{tabular}} & \textbf{\begin{tabular}[c]{@{}c@{}}ROI(\%)\\ Nasdaq\end{tabular}} & \textbf{\begin{tabular}[c]{@{}c@{}}ROI(\%)\\ S\&P500\end{tabular}} \\ \midrule
\textit{\textbf{Test Set 1}} & 30 Days & 26.98 & 3.28 & -6.48 & 4.625 & 2.88 \\
\textit{\textbf{}} & 60 Days & 30.33 & -6.98 & -21.88 & -8.608 & -8.24 \\
\textit{\textbf{}} & 90 Days & 26.23 & 16.908 & -15.45 & -7.403 & -4.94 \\ \midrule
\textit{\textbf{Test Set 2}} & 30 Days & 19.77 & 1.509 & -2.08 & 2.212 & 0.739 \\
\textit{\textbf{}} & 60 Days & 36.91 & 0.38 & -2.28 & 4.249 & 0.829 \\
\textit{\textbf{}} & 90 Days & 79.57 & 12.79 & 0.467 & 7.844 & 5.14 \\ \midrule
\textit{\textbf{Test Set 3}} & 30 Days & 20.85 & 1.814 & -0.807 & -0.281 & -0.101 \\
\textit{\textbf{}} & 60 Days & 53.16 & 4.17 & 0.58 & 3.087 & 2.521 \\
\textit{\textbf{}} & 90 Days & 83.97 & 20.20 & 4.539 & 7.70 & 6.029 \\ \midrule
\textit{\textbf{Test Set 4}} & 30 Days & 8.5 & 4.30 & 3.55 & 5.280 & 4.675 \\
\textit{\textbf{}} & 60 Days & 50.78 & -0.78 & -4.54 & 0.923 & 0.013 \\
\textit{\textbf{}} & 90 Days & 60.81 & 4.71 & 0.619 & 4.866 & 4.837 \\ \midrule
\textit{\textbf{Test set 5}} & 30 Days & 37.66 & 2.97 & 2.616 & 5.772 & 4.417 \\
\textit{\textbf{}} & 60 Days & 63.53 & 5.095 & 1.584 & 12.478 & 7.797 \\
\textit{\textbf{}} & 90 Days & 55.36 & -32.86 & -35.018 & -18.722 & -27.522 \\ \bottomrule
\end{tabular}
\end{table}
In order to further assess the performance of the proposed framework, the return on investment of DeepBreath was compared with four benchmark portfolios namely the Dow Jones Industrial (DJI), the Euro Stoxx $50$ ETF, Nasdaq, and S\&P$500$ \citep{serletis2009mean}. Our results are reported in Table~\ref{tab:RoI}. For instance, for test set 5, our framework achieved a $37.66\%$, $63.53\%$, and $55.36\%$ return on investment (ROI) over thirty, sixty, and ninety days, respectively. Meanwhile, for the same investment periods, the ROI was $2.97\%$, $5.095\%$, and $-32.86\%$ for DJI, and $2.616\%$, $1.584\%$, $-35.018\%$ for Euro Stoxx 50 ETF, $5.772\%$, $12.478\%$, $-18.722\%$ for Nasdaq and finally $4.417\%$, $7.797\%$, $-27.522\%$ for S\&P500.
\section{Discussion}\label{sec:Discussion}
DeepPocket was evaluated against five real-life datasets over three distinct investment periods including during the Covid-19 crisis, for which it clearly outperformed market indexes. The results obtained during the Covid-19 crisis are particularly remarkable as DeepPocket managed to generate a profit by exploiting profitable assets, such as Amazon and Facebook, in addition to exploiting offline knowledge such as what it learned from the 2008 global financial recession. This performance is in contrast with market indexes, which were collapsing during the same period.  
The reason is that our deep neural network learns the underlying rules, connections and predictable patterns directly from the data.  The network adapts as new information becomes available, which allows the system to make better predictions based on insight acquired from previously analyzed information.  The amount of complex information that it can process far surpasses largely human capacities.  The network may detect anomalies such as unexpected spikes, drops, level shifts and trend changes.  In contrast to traders, DeepPocket is not prone to such a high degree of personal biases, conflicts of interest, and emotional decisions.  Indeed, as reported recently in \footnote{https://www.bloomberg.com/news/articles/2020-02-11/robot-analysts-outwit-humans-in-study-of-profit-from-stock-calls}, robot-analysis systems, like DeepPocket, tend to outperform traditional investment strategies.

\subsection{Conclusion}\label{subsec:conclusion}
In this paper, we proposed DeepPocket, a portfolio management framework that aims to increase the expected return on investment while hedging the risk. This framework exploits the time-varying correlations between financial instruments. These correlations are represented in a graph whose nodes correspond to the financial instruments and whose edges correspond to pair-wise correlation functions between assets. DeepPocket consists of a restricted, stacked autoencoder (RSAE) for feature extraction, and an actor--critic reinforcement learning agent. The actor--critic framework consists of two convolutional networks in which the actor learns and enforces an investment policy, and the critic, in turn, evaluates the policy to establish the best course of action by constantly reallocating the various portfolio assets to maximize the expected return on investment. The agent is initially trained offline with online stochastic batching on historical data. As new data becomes available, the agent is trained online with a passive concept-drift approach to handle unexpected changes in the underlying data distribution.  

The approach has many benefits for practitioners such as seeking investment opportunities, identifying market patterns, evaluating trading strategies, and automating workflows, just to mention a few.  By finding predictable patterns, structures and trends, the system may provide valuable recommendations and insight to traders.  Risk modelling and forecasting, as well as their accuracy, may be improved using the insight acquired from the data.  Traditional risk models have demonstrated their ineffectiveness in coping with global financial crisis.  As shown by our results, DeepPocket successfully managed such risks during the Covid-19 crisis: a characteristic which has direct repercussions for portfolio risk management and compliance.  

DeepPocket assumes a liquid market which means that many buyers and sellers are available, while price variations remain relatively small.  We plan to target this problem in a future work, in order to address situations in which buyers and sellers are in short supply.  Besides portfolio allocation, we plan to have DeepPocket select the stocks directly in addition to developing deep learning models more suitable for long-term investment strategies.

\bibliographystyle{elsevierbibstyle}\biboptions{authoryear}
\bibliography{Reference}
\end{document}